\documentclass[acmsmall,nonacm]{acmart}

\usepackage{multirow}
\usepackage{array}
\usepackage{enumitem}
\usepackage[ruled,vlined,linesnumbered]{algorithm2e}
\usepackage[capitalize,noabbrev]{cleveref}
\usepackage{placeins}

\newtheorem{proposition}{Proposition}

\begin{document}

\title[Beyond Semantic Relevance: Evidence-Supported Stopping Depth]{Beyond Semantic Relevance: Evidence-Supported Stopping Depth in Open-Set Hierarchical Event Linking}

\author{Shengqi Zhou}
\orcid{0009-0004-2451-3422}
\affiliation{\institution{Northeast Petroleum University}\city{Daqing}\state{Heilongjiang}\country{China}}
\email{shbiao@stu.nepu.edu.cn}
\author{Zhigang Liu}
\affiliation{\institution{Northeast Petroleum University}\city{Daqing}\state{Heilongjiang}\country{China}}
\author{Wenxiang Zhu}
\affiliation{\institution{Tongji University}\city{Shanghai}\country{China}}
\author{Boyong Du}
\affiliation{\institution{Chongqing University}\city{Chongqing}\country{China}}
\author{Zhentao Cang}
\affiliation{\institution{Northwestern Polytechnical University}\city{Xi'an}\state{Shaanxi}\country{China}}

\begin{abstract}
Linking open-world social posts and news headlines to hierarchical event ontologies requires deciding not only topical relevance, but also \emph{evidence-supported stopping depth}: when textual evidence justifies descending to a fine-grained leaf versus backing off to a coarser ancestor.
Across two platforms (X/Twitter and Hacker News), nine Storylines, and 1,177 annotated records (including a prospective blind split, \texttt{Reserve-160}, with 80 dual-audited items, $\kappa=0.852$), we observe an \emph{All-Backoff Paradox}: 66.4\% (184/277) of held-out errors stem from over-descending into unsupported Atomic leaves, so blanket backoff raises exact match by $+3.62$--$16.25$~pp on low-density splits yet lowers high-density accuracy by $-7.50$~pp.
Benchmarking 15+ IR, NLI, selective-prediction, and hierarchical-hedging (DARTS) baselines shows none reliably beats All-Backoff.
Through causal falsifications, counterfactual span masking, and direct $\Delta(e,a)$ verification on \texttt{Qwen2.5-7B} and \texttt{Phi-4-mini}, we link 3-null representation centering ($L_{\mathrm{cal}}$) and prompt log-odds via card-intrinsic \emph{Granularity Gap} $\Delta(e,a)$.
We unify scope gating, path decoding, and zero-parameter calibration (\textbf{\texttt{AdaDepth}}) inside \textbf{\texttt{EviLink}}.
Direct $\Delta(e,a)$ verification excels on high-gap Hacker News (0.87--0.90~AUC with auto-extracted $\widehat{\Delta}_{\mathrm{auto}}$; 0.88--0.95 with manual $\Delta$) yet falls below All-Backoff on noisy X/Twitter ($-1.61$~pp), whereas \texttt{AdaDepth} achieves the top $+2.97$~pp 8-cell mean gain (7/8 cells) and stronger Risk--Coverage frontiers.
\end{abstract}



\maketitle

\section{Introduction}\label{sec:intro}
Real-world information systems routinely organize dynamic public discourse (from breaking news headlines to informal social media streams) into multi-level event knowledge graphs spanning broad \emph{Storylines}, intermediate \emph{Episodes}, and fine-grained \emph{Atomic Events} \citep{hogan2021,yu2023event}.
To populate such hierarchies reliably under open-world streams, an automated linker must solve two distinct decision problems.
First, it must identify \emph{semantic relevance} and open-set membership: does a post belong to a registered event lineage at all, or should it be rejected as out-of-scope (\texttt{none})?
Second, and far less understood in information retrieval (IR), it must determine the \textbf{evidence-supported stopping depth}: \emph{when semantic relevance alone is insufficient to substantiate a fine-grained leaf claim, how can an information system detect the boundary of evidential sufficiency and decide when to stop descending?}

\begin{figure}[t]
\centering
\includegraphics[width=\textwidth]{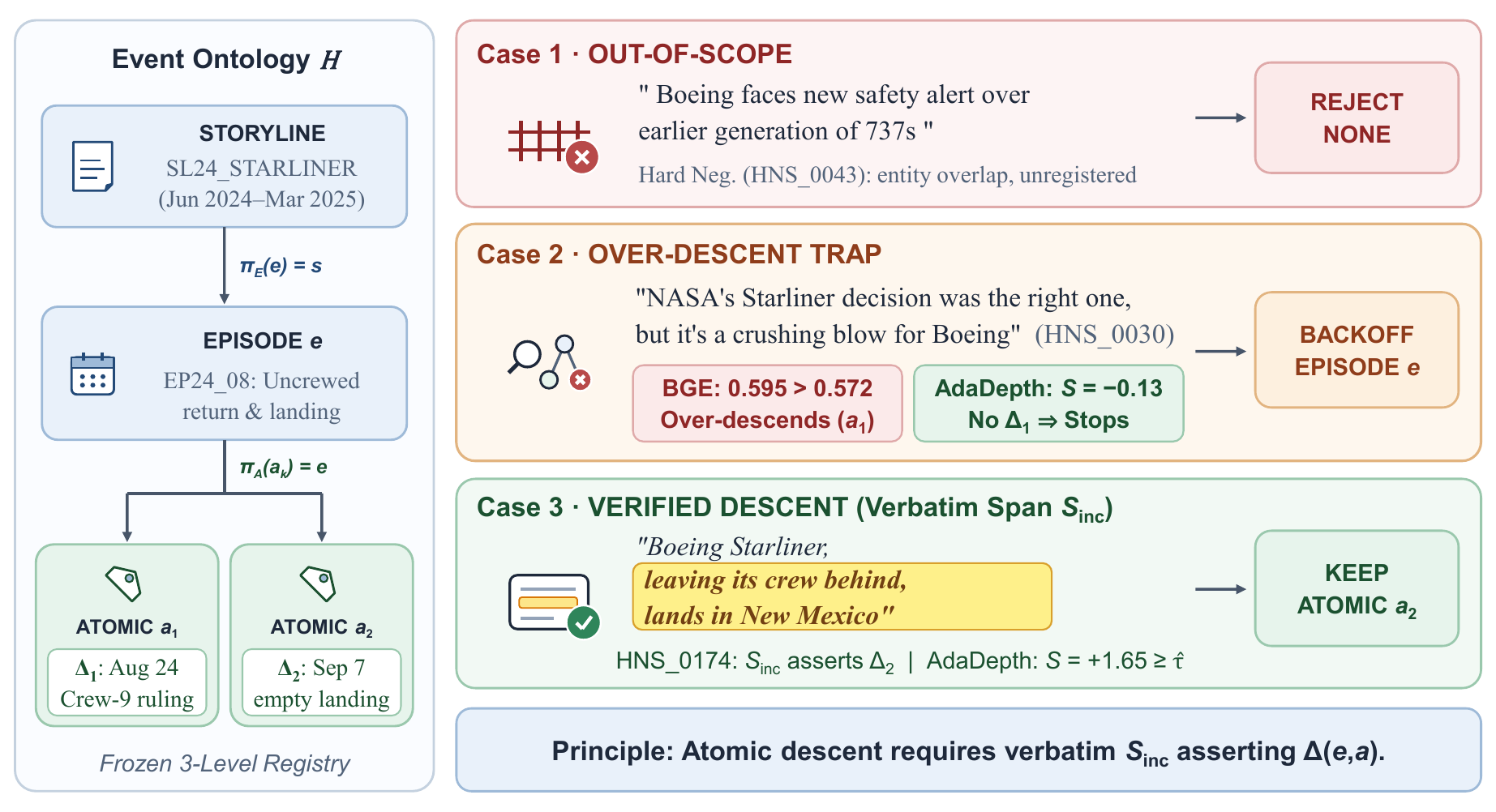}
\caption{Motivating task illustration using verbatim records from our frozen Hacker News benchmark (\texttt{SL24\_STARLINER\_CFT}). \textbf{Left:}~A frozen three-level event registry ontology $\mathcal{H}$ pairs each Atomic leaf $a_k$ with an explicit Incremental Assertion $\Delta(e,a_k)$ beyond its parent Episode $e$. \textbf{Right:}~Three evidential stopping-depth decisions: Case~1 (\texttt{HNS\_0043}) shares entities (``Boeing'', ``safety'') but lies outside the registry (\texttt{NONE}); Case~2 (\texttt{HNS\_0030}) exhibits higher dense cosine similarity to leaf $a_1$ than parent $e$ (\texttt{BGE-large}: $0.595 > 0.572$), luring relevance-based systems into the \emph{over-descending trap}, whereas \texttt{AdaDepth} stops at \texttt{episode\_only}; Case~3 (\texttt{HNS\_0174}) contains a verbatim span $S_{\mathrm{inc}}(p_3)$ asserting $\Delta(e,a_2)$, licensing verified descent to \texttt{atomic\_exact}. Displayed \texttt{AdaDepth} scores are frozen \texttt{Qwen2.5-7B} \texttt{AdaDepth v1.0} outputs from the released artifact.}
\label{fig:task}
\Description{Two-panel motivating figure showing the Starliner event ontology tree on the left and three real Hacker News decision cases on the right: open-set rejection, the over-descending trap where BGE-large fails and AdaDepth backs off, and verified atomic descent with a highlighted verbatim evidence span.}
\end{figure}

\Cref{fig:task} illustrates why semantic relevance and evidential stopping depth diverge sharply in real discourse.
Consider the registered Episode \texttt{EP24\_08} covering NASA's August--September 2024 decision to return Boeing's Starliner capsule uncrewed and its subsequent autonomous touchdown.
The headline \texttt{HNS\_0030} (\emph{``NASA's Starliner decision was the right one, but it's a crushing blow for Boeing''}) is undeniably relevant to both the Episode and its Atomic child $a_1$ (the August~24 Crew-9 reassignment ruling).
Indeed, a modern dense bi-encoder (\texttt{BAAI/bge-large-en-v1.5}; \citealp{xiao2024cpack}) scores the leaf higher than the parent ($\mathrm{sim}(p_2,a_1)=0.595 > \mathrm{sim}(p_2,e)=0.572$), and an unconstrained large language model (LLM) descends directly to \texttt{atomic\_exact}.
Yet the headline is broad evaluative commentary: it never asserts the Crew-9 reassignment or formal August~24 ruling required by the incremental assertion $\Delta(e,a_1)$.
Under strict evidence grounding, the only defensible link is \texttt{episode\_only}.
Conversely, \texttt{HNS\_0174} (\emph{``Boeing Starliner, leaving its crew behind, lands in New Mexico''}) contains a verbatim span $S_{\mathrm{inc}}(p_3)$ explicitly asserting the uncrewed landing $\Delta(e,a_2)$, warranting full descent to \texttt{atomic\_exact}.

When we formalize this problem inside our unified three-stage open-set hierarchical event linking framework (\textbf{\texttt{EviLink}}) and evaluate across two contrasting platforms, noisy X/Twitter posts (\texttt{Pilot-200} and held-out \texttt{Expansion-497}) and concise Hacker News story headlines (\texttt{HN-Stories-320} and prospective \texttt{Reserve-160}, preregistered and frozen prior to Gold annotation), we observe an empirical phenomenon that we term the \textbf{All-Backoff Paradox}.
Although a scope-gated LLM linker reaches 98.8\% Storyline accuracy and 86.2\% Episode accuracy on held-out X/Twitter Expansion, its end-to-end Hierarchical Exact Match (HEM) plateaus at 0.443 because \textbf{66.4\% (184/277) of all errors are over-descending errors} into unsupported Atomic leaves.
Blindly reverting every Atomic prediction to its parent Episode (\emph{All-Backoff}) raises Expansion HEM from 0.443 to 0.479 ($+3.62$~pp) and lifts blind HN Stories HEM from 0.366 to 0.497 ($+13.13$~pp), yet the same blanket policy lowers high-density Pilot-200 HEM from 0.750 to 0.675 ($-7.50$~pp).
The optimal stopping depth is thus distribution-dependent: neither unconditional descent nor unconditional backoff provides a universally reliable operating policy.

Can modern IR retrievers, neural rerankers, NLI entailment models, or selective-prediction uncertainty estimators resolve this boundary?
Our benchmark across 15+ competitive baselines shows that \emph{semantic relevance is mismatched to stopping-depth calibration}:
\begin{itemize}[leftmargin=*,nosep]
  \item \textbf{Neural NLI suffers severe inversion and domain fragility.} \texttt{DeBERTa-v3-large} fine-tuned on five NLI corpora \citep{he2023debertav3,laurer2024less} collapses to \textbf{0.304~AUC} (below random inversion) on held-out X/Twitter Expansion and drops HEM by $-8.25$~pp below All-Backoff ($p=2\times 10^{-6}$), while \texttt{BART-large-MNLI} \citep{lewis2020bart} hovers near chance (0.522~AUC) and loses to All-Backoff on HN Stories.
  \item \textbf{Dense bi-encoders and cross-encoder rerankers over-descend on lexical and topical overlap.} \texttt{BGE-large} retains 136/138 over-descending errors on X/Twitter (0.435~HEM, $-4.43$~pp vs.\ All-Backoff) and 80/97 on HN Stories (0.406~HEM, $-9.06$~pp), while \texttt{MiniLM-L6} cross-encoder \citep{wang2020minilm} underperforms All-Backoff on both held-out benchmarks ($-4.43$~pp and $-3.12$~pp).
  \item \textbf{Standard selective-prediction and hierarchical hedging baselines fail to generalize across base-rate shifts.} Maximum softmax probability (MSP), negative entropy \citep{geifman2017selective}, and hierarchical accuracy--specificity hedging (DARTS; \citealp{deng2012hedging}) either preserve severe over-descending (MSP: 0.384~HEM on HN, $-11.25$~pp vs.\ All-Backoff) or fail to beat All-Backoff on any held-out benchmark (DARTS: $+0.00$~pp on X-Exp, $-1.25$~pp on HN).
\end{itemize}

To understand \emph{why} stopping-depth signals behave differently across genres and models, we run three controlled causal falsifications and counterfactual token-masking audits across two open-weight LLM families (\texttt{Qwen2.5-7B-Instruct} and \texttt{Phi-4-mini-instruct}).
We falsify free-form LLM self-auditing (0.495~AUC), falsify prompt-visible sibling-list attribution across all three datasets (hiding sibling cards never degrades and often improves discrimination), and find a sharp architectural divergence in content-free contextual calibration \citep{zhao2021calibrate}: subtracting 3-null baselines severely degrades instruction-tuned prompt logits (e.g., $-0.178$ to $-0.213$~AUC) yet reliably improves final-layer hidden-state margins ($L_{\mathrm{cal}}$) in 4/4 model--platform cells while leaving within-pair rank order unchanged.
Most importantly, the regime inversion between X/Twitter (where $L_{\mathrm{cal}}$ leads) and HN Stories (where prompt log-odds \texttt{SymDiff} and \texttt{Direct\_AB} reach 0.79--0.89~AUC) follows the card-intrinsic \textbf{Granularity Gap} $\Delta(e,a)$, and is confirmed mechanistically via matched-$k$ counterfactual masking of human-annotated verbatim spans $S_{\mathrm{inc}}(p)$ ($p=0.00035$ on \texttt{Qwen2.5-7B}; $p=0.00086$ on \texttt{Phi-4-mini}).

Translating this discovery into a deployable IR architecture, we complete \textbf{\texttt{EviLink}} with \textbf{\texttt{AdaDepth}} (Stage~3), a zero-parameter, zero-target-label stopping-depth calibrator operating solely over standard parent--child Event Card descriptions $(c_e,c_a)$ in $\mathcal{H}$ (requiring no manually authored $\Delta(e,a)$ text at inference time).
\texttt{AdaDepth} combines an offline ontology-intrinsic gap gate $w_{\mathrm{gap}}(e,a)\in[0,1]$, robust MAD-standardized representation and prompt sufficiency scores, and a \textbf{Null-Anchored Prevalence Threshold} $\hat{\tau}_{\mathrm{Ada}} = \mathrm{Quantile}_{1-\hat{\pi}_L}(S_{\mathrm{AdaDepth}})$ derived strictly from unlabeled 3-null centering.
Across a $6\times$ shift in true Atomic base rates (from 10.7--11.1\% on HN Stories to 65.7\% on X-Pilot), \texttt{AdaDepth} tracks target prevalence automatically (2.8~pp MAE on \texttt{Phi-4-mini}), eliminates cross-model prompt logit drift (lifting \texttt{Phi-4-mini} \texttt{Direct\_AB} from 0.369 to 0.500~HEM on HN-CP1-320 and from 0.419 to 0.588~HEM on prospective \texttt{Reserve-160}), and achieves the highest 8-cell mean gain over All-Backoff ($+2.97$~pp, winning 7/8 benchmark cells and 2/2 \texttt{Reserve-Dual-80} cells) under both inclusive and strict exact-$K$ matched-budget protocols.
Moreover, a dedicated direct $\Delta(e,a)$ verification baseline confirms both sides of our thesis: directly verifying auto-extracted $\widehat{\Delta}_{\mathrm{auto}}(e,a)$ reaches 0.87--0.90~AUC on crisp high-gap HN Stories (matching manual $\Delta(e,a)$ on prospective \texttt{Reserve-160} at 0.588~HEM) yet drops below All-Backoff on noisy low-gap X/Twitter ($-1.61$~pp), explaining why gap-weighted fusion is needed for cross-regime deployment.

Together, \textbf{\texttt{EviLink}} unifies open-set scope gating (Stage~1), legal path decoding (Stage~2), and zero-parameter stopping-depth calibration via \texttt{AdaDepth} (Stage~3) within an IR decision-theoretic evaluation framework.
This paper makes four principal contributions:
\begin{enumerate}[leftmargin=*,nosep]
  \item \textbf{Formalization and IR Empirical Diagnosis (Beyond Semantic Relevance):} We formalize evidence-supported stopping depth via card-level Incremental Assertions $\Delta(e,a)$ and verbatim evidence spans $S_{\mathrm{inc}}(p)$, identify the All-Backoff Paradox, and demonstrate across 15+ modern IR, reranker, NLI, selective-prediction, and hierarchical-hedging baselines that topical relevance and standard entailment cannot govern fine-grained stopping depth.
  \item \textbf{Unified \texttt{EviLink} Framework and Zero-Parameter \texttt{AdaDepth} Calibrator:} We design the three-stage \texttt{EviLink} architecture with \texttt{AdaDepth}, prove exact within-pair card-bias invariance and decoupled prevalence-anchored quantile control (Propositions~1--2), and isolate unlabeled prevalence anchoring ($\hat{\pi}_L$) from Granularity-Gap-aware score fusion ($S_{\mathrm{AdaDepth}}$) in a full cross-ablation, achieving the top $+2.97$~pp mean gain over All-Backoff across all 8 benchmark cells.
  \item \textbf{Two-Platform Nine-Storyline Benchmark and Stage-2 Blind Validation:} We construct and evaluate \textbf{1,177 formally annotated records} (640 dual-audited) across X/Twitter and Hacker News, including a two-stage preregistered blind protocol whose Stage-2 prospective \texttt{Reserve-160} predictions (12 artifacts, 72 leaf candidates) were locked prior to dual human annotation ($\kappa=0.852$) and verified for pre-annotation consistency before scoring.
  \item \textbf{Mechanistic Causal Evidence:} Through 3-dataset causal sibling ablations, 3-null architectural divergence analysis, Granularity-Gap stratification, and matched-$k$ counterfactual span masking, we establish when and why representation margins versus prompt log-odds track incremental textual evidence.
\end{enumerate}

\section{Related Work}\label{sec:related}

\subsection{Event coreference, event linking, and hierarchical classification}
Cross-document event coreference clusters co-referring event mentions across news articles using lexical, argument, and discourse-coherence features \citep{bagga1999,bejan2010,lee2012,cybulska2014,lu2018,cattan2021,chen2023discourse,chen2025coherence}, while social event detection, LLM clustering, and LLM-based event extraction frameworks organize open-world streams into structured event schemas and graphs \citep{gao2024cdee,li2024relational,chu2025llmaec,li2025ee_llm}.
Closer to our setting, \citet{yu2023event} formalize flat \emph{Event Linking} to Wikipedia event pages \citep{reimers2019,nogueira2019}, \citet{ou2023hierarchical} introduce \emph{Hierarchical Event Grounding} to align text mentions to multi-level event schemas, and \citet{hoveyda2024conversational} show in conversational entity linking over informal social discussions that zero-shot linkers degrade sharply when transferred from encyclopedic benchmarks to sparse, domain-specific knowledge bases.
Our formulation diverges from Hierarchical Event Grounding and flat event linking along three concrete axes: (i)~\emph{open-set streams}, where candidate posts frequently fall outside the registered event ontology and must be rejected as \texttt{none} \citep{scheirer2013,bendale2016}; (ii)~\emph{explicit evidential criteria}, requiring a verbatim textual span $S_{\mathrm{inc}}(p)$ asserting the parent--child incremental claim $\Delta(e,a)$ rather than topical or schema-type similarity alone; and (iii)~\emph{deepest legal-path output}, where an unsupported leaf assignment must back off cleanly to a valid ancestor $(s,e,\varnothing)$ inside the taxonomy \citep{silla2011,hogan2021}.
Although retrieval-augmented generation, LLM-based IR pipelines, and knowledge-graph prompting \citep{lewis2020,brown2020,wei2022,zhao2024kgcot,baghdasaryan2024knowledge,zhu2025llmir} improve coarse Storyline routing, over two-thirds of residual errors occur when linkers over-descend past the deepest ancestor supported by the text.

\subsection{Hierarchical hedging, selective prediction, and IR calibration}
Choosing how deep to predict in a taxonomy relates directly to accuracy--specificity hedging in visual hierarchies (\textbf{DARTS}; \citealp{deng2012hedging}), selective classification with a reject option \citep{elyaniv2010selective,geifman2017selective,franc2023reject}, and rigorous IR evaluation under replicate splits and statistical reliability \citep{voorhees2017replicates}, evaluated via \emph{Risk--Coverage} curves, \textbf{AURC}, and \textbf{E-AURC} \citep{geifman2019bias}.
In IR and neural ranking, calibration and uncertainty studies have similarly highlighted that neural rankers and LLM-based retrieval systems produce miscalibrated score scales that degrade risk-sensitive decisions \citep{penha2021calibration,yan2022scale,zhu2025llmir}, while recent work on uncertainty quantification for retrieval-augmented reasoning \citep{soudani2026uncertainty}, long-document abstention \citep{buchmann2024abstain}, and distribution-free Conformal Risk Control \citep{angelopoulos2024conformal} seeks calibrated abstention or error-rate guarantees.
However, \emph{evidence-supported stopping depth} differs from traditional hierarchical hedging and conformal control in three respects: (i)~traditional hedging (e.g., DARTS) presupposes well-calibrated posterior probabilities over candidate leaves and maximizes expected utility under a static accuracy--specificity trade-off, whereas open-world event linkers face severe probability miscalibration and must verify textual support for an explicit incremental proposition $\Delta(e,a)$; (ii)~unlike binary abstention that discards rejected queries, backing off from an unsupported Atomic leaf preserves a valid coarser Episode link $(s,e,\varnothing)$; and (iii)~unlike Conformal Risk Control, which requires a labeled held-out calibration set from the target distribution, true leaf-support base rates shift by more than $6\times$ across storylines and platforms, breaking static thresholds, MSP confidence, and uncalibrated DARTS posteriors.

\subsection{Semantic retrieval, NLI, label-shift calibration, and faithfulness}
Dense bi-encoders (\texttt{BGE-large}; \citealp{xiao2024cpack}), cross-encoder rerankers (\texttt{MiniLM-L6}; \citealp{wang2020minilm}), lexical \texttt{BM25} \citep{robertson2009}, and zero-shot NLI classifiers (\texttt{DeBERTa-v3-large} \citealp{he2023debertav3,laurer2024less}; \texttt{BART-large-MNLI} \citealp{lewis2020bart}) form the backbone of modern IR verification pipelines.
Yet topical similarity is symmetric and additive with shared entities, while NLI classifiers trained on premise--hypothesis pairs frequently confuse topical implication with strict incremental event claims.
Estimating shifting class priors without target labels has a rich history in \emph{label-shift adaptation}, including EM prior adjustment \citep{saerens2002adjusting}, Black Box Shift Estimation (\textbf{BBSE}; \citealp{lipton2018bbse}), and bias-corrected maximum likelihood (\textbf{MLLS}; \citealp{alexandari2020mlls}).
In contrast to our setting, classical EM, BBSE, and MLLS all require a labeled source validation split to estimate the classifier's confusion matrix or temperature-scaled source posteriors, which fails when deploying onto an unseen event ontology with zero labeled examples.
Drawing instead on content-free contextual calibration \citep{zhao2021calibrate} and surface-form competition analysis \citep{holtzman2021surface}, we expose a sharp boundary condition in modern instruction-tuned LLMs: 3-null subtraction breaks prompt logits because null inputs trigger refusal shifts, yet succeeds consistently in final-layer hidden-state space ($L_{\mathrm{cal}}$) and provides a fully unlabeled target prevalence anchor $\hat{\pi}_L$ requiring no source validation labels.
Finally, building on counterfactual rationale faithfulness (\textbf{ERASER}; \citealp{deyoung2020eraser}), we introduce matched-$k$ $\Delta$-token masking to test counterfactual evidence sensitivity and verify that stopping-depth signals respond faithfully to human-annotated incremental evidence spans rather than spurious entity cues.

\section{Problem Formulation: Evidence-Supported Stopping Depth}\label{sec:problem}

\subsection{Frozen hierarchical event ontology and legal output matrix}
Let a frozen three-level event registry ontology be $\mathcal{H} = (\mathcal{S}, \mathcal{E}, \mathcal{A}, \pi_E, \pi_A)$, where $\mathcal{S}$ is the set of registered Storylines, $\mathcal{E}$ is the set of Episodes with parent map $\pi_E: \mathcal{E}\to\mathcal{S}$, and $\mathcal{A}$ is the set of active Atomic Events with parent map $\pi_A: \mathcal{A}\to\mathcal{E}$.
For each Episode $e\in\mathcal{E}$, denote its active Atomic children by $\mathcal{A}_e = \{a\in\mathcal{A}:\pi_A(a)=e\}$.
Each node $v\in\mathcal{S}\cup\mathcal{E}\cup\mathcal{A}$ is represented by a structured textual \emph{Event Card} $c_v$ specifying its canonical title, temporal window, and scope boundaries.

Given an incoming post or headline $p$ (text, timestamp, and optional media indicator), an open-set hierarchical event linker outputs a legal tuple $y = (s, e, a, \ell) \in \widetilde{\mathcal{Y}}$ whose assignment level $\ell \in \{\texttt{atomic\_exact}, \texttt{episode\_only}, \texttt{storyline\_only}, \texttt{none}\}$ satisfies the deterministic legality constraints:
\begin{equation}\label{eq:legal_matrix}
\begin{aligned}
\texttt{atomic\_exact}:&\quad s\in\mathcal{S},\ e\in\mathcal{E},\ a\in\mathcal{A},\ \pi_E(e)=s,\ \pi_A(a)=e;\\
\texttt{episode\_only}:&\quad s\in\mathcal{S},\ e\in\mathcal{E},\ a=\varnothing,\ \pi_E(e)=s;\\
\texttt{storyline\_only}:&\quad s\in\mathcal{S},\ e=a=\varnothing;\\
\texttt{none}:&\quad s=e=a=\mathrm{NONE}.
\end{aligned}
\end{equation}
A deterministic validator $V_{\mathcal{H}}: \widetilde{\mathcal{Y}} \to \widetilde{\mathcal{Y}} \times \{\mathrm{valid}, \mathrm{invalid}\}$ enforces registry membership, parent-path integrity, and level-specific field filling without altering semantic choices, guaranteeing zero database schema violations.

\subsection{Incremental assertions \texorpdfstring{$\Delta(e,a)$}{Delta(e,a)}, verbatim spans \texorpdfstring{$S_{\mathrm{inc}}(p)$}{S_inc(p)}, and Granularity Gap}
Why does topical relevance fail at the Episode--Atomic boundary?
Because a child Event Card $c_a$ shares almost all primary entities, locations, and time anchors with its parent Episode Card $c_e$; any post discussing $e$ is automatically topically relevant to $a$.
We formalize the evidential sufficiency boundary via two coupled constructs:
\begin{itemize}[leftmargin=*,nosep]
  \item \textbf{Card-Level Incremental Assertion $\Delta(e,a)$:} For every parent--child pair $(e,a)$ with $\pi_A(a)=e$, $\Delta(e,a)$ specifies the minimal non-background propositional claim beyond Episode $e$ required to license descent to leaf $a$, paired with explicit \emph{disqualifying parent framings} (e.g., general reaction, retrospective commentary, or multi-aspect impact).
  \item \textbf{Post-Level Verbatim Evidence Span $S_{\mathrm{inc}}(p)$:} Post $p$ is licensed to receive \texttt{atomic\_exact} for leaf $a$ if and only if $p$ contains a verifiable verbatim substring $S_{\mathrm{inc}}(p) \subseteq p$ asserting $\Delta(e,a)$ as the post's central proposition rather than incidental background.
\end{itemize}
Each pair $(e,a)$ also exhibits an ontology-intrinsic \textbf{Granularity Gap} measuring how sharply $\Delta(e,a)$ narrows the parent Episode scope, represented both as an expert pre-Gold ordinal tag $g(e,a)\in\{\texttt{low},\texttt{medium},\texttt{high}\}$ and as a computable card-intrinsic scalar $w_{\mathrm{gap}}(e,a)\in[0,1]$ defined in \S\ref{sec:method}.

\subsection{Post-proposal stopping depth, Risk--Coverage, and Accuracy--Specificity}
Given any upstream open-set linker that emits a leaf proposal set
\begin{equation}\label{eq:cleaf}
\mathcal{C}_{\mathrm{leaf}} = \bigl\{(p_i, \hat{s}_i, \hat{e}_i, \hat{a}_i) : \hat{\ell}_i = \texttt{atomic\_exact}\bigr\},
\end{equation}
post-proposal stopping-depth calibration assigns each candidate a scalar evidential sufficiency score $S_i = S(p_i; \hat{e}_i, \hat{a}_i)\in\mathbb{R}$ and applies a legal depth decision at operating threshold $\tau$:
\begin{equation}\label{eq:decision}
\mathrm{Dec}_{\tau}(p_i) =
\begin{cases}
(\hat{s}_i, \hat{e}_i, \hat{a}_i, \texttt{atomic\_exact}), & S_i \ge \tau \quad\text{(KEEP)},\\
(\hat{s}_i, \hat{e}_i, \varnothing, \texttt{episode\_only}), & S_i < \tau \quad\text{(BACKOFF)}.
\end{cases}
\end{equation}
Each leaf candidate falls into one of three disjoint outcome partitions: \emph{Keep-Correct} ($\mathrm{KeepOK}_i = 1$ when $y_i$ matches $(\hat{s}_i,\hat{e}_i,\hat{a}_i)$ at \texttt{atomic\_exact}), \emph{Rescuable Over-Descend} ($\mathrm{BackoffOK}_i = 1$ when $y_i$ matches $(\hat{s}_i,\hat{e}_i,\varnothing)$ at \texttt{episode\_only}), or \emph{Upstream Error} (wrong Episode/Storyline or Gold \texttt{none}, where neither KEEP nor BACKOFF achieves exact match).
Candidates with $\mathrm{KeepOK}_i = 1 \lor \mathrm{BackoffOK}_i = 1$ are \emph{decisive}, and $\mathrm{OD}_{\mathrm{raw}}$ counts all leaf candidates whose Gold level is \texttt{episode\_only}.

To evaluate stopping-depth systems rigorously from an IR decision-making standpoint, we combine four complementary metric families under frozen \textbf{Protocol v2.1}:
\begin{enumerate}[leftmargin=*,nosep]
  \item \textbf{End-to-End Accuracy:} Hierarchical Exact Match $\mathrm{HEM} = \frac{1}{N}\sum_{i=1}^N \mathbb{I}[\hat{y}_i = y_i]$, evaluated strictly at each method's deployable zero-label operating point (never oracle-tuned on test labels), alongside retained Atomic links ($\mathrm{Ret}$) and surviving over-descending errors ($\mathrm{OD}$).
  \item \textbf{Threshold-Free Ranking:} Decisive ROC-AUC ($\mathrm{AllAUC}$), weighted within-card-pair concordance ($\mathrm{InPairAUC}$, isolating post discrimination from card-pair offset bias), and Atomic retention under fixed over-descending budgets $B_q = \lfloor q\cdot \mathrm{OD}_{\mathrm{raw}}\rfloor$ ($q\in\{0.25, 0.50\}$; $\mathrm{Ret@25\%}$, $\mathrm{Ret@50\%}$) with conservative block tie-handling.
  \item \textbf{Selective-Prediction Risk--Coverage (AURC / E-AURC):} Ordering $\mathcal{C}_{\mathrm{leaf}}$ by decreasing $S_i$, at leaf-keep coverage $c = k/|\mathcal{C}_{\mathrm{leaf}}| \in (0,1]$, the selective leaf risk is $R(c) = \frac{1}{k}\sum_{j=1}^k \mathbb{I}[\mathrm{KeepOK}_{\pi(j)}=0]$ \citep{elyaniv2010selective,geifman2017selective}. $\mathrm{AURC} = \frac{1}{|\mathcal{C}_{\mathrm{leaf}}|}\sum_{k=1}^{|\mathcal{C}_{\mathrm{leaf}}|} R(k/|\mathcal{C}_{\mathrm{leaf}}|)$ measures aggregate risk across all operating coverages, and $\mathrm{E\text{-}AURC} = \mathrm{AURC} - \mathrm{AURC}_{\mathrm{oracle}}$ isolates excess risk above the optimal ranker \citep{geifman2019bias}.
  \item \textbf{Accuracy--Specificity Frontier:} Following \citet{deng2012hedging}, sweeping $\tau$ traces exact-match accuracy $\mathrm{HEM}(\tau)$ against mean output ontology depth $\bar{D}(\tau) = 2 + \frac{1}{|\mathcal{C}_{\mathrm{leaf}}|}\sum_{i\in\mathcal{C}_{\mathrm{leaf}}}\mathbb{I}[S_i \ge \tau] \in [2, 3]$ (depth~2 = \texttt{episode\_only}, depth~3 = \texttt{atomic\_exact}), showing whether a scorer achieves higher accuracy at equal or greater specificity.
\end{enumerate}

\section{Two-Platform, Nine-Storyline Benchmark and Stage-2 Blind Protocol}\label{sec:data}
To guard against single-platform artifacts and retrospective leakage, our evaluation spans two contrasting genres, nine Storylines, and \textbf{1,177 formally annotated records}, including a two-stage preregistered Hacker News blind benchmark (\Cref{tab:resources}).

\begin{table}[t]
\centering
\caption{Overview of the two-platform, nine-Storyline frozen benchmark suite (1,177 formally annotated Gold records across Platform~1 and two-stage blind Platform~2).}
\label{tab:resources}
\footnotesize
\setlength{\tabcolsep}{3.6pt}
\resizebox{\textwidth}{!}{%
\begin{tabular}{llcccccccccc}
\toprule
Dataset & Platform / Status & Storylines & Ep. & Atomics & $N$ & $|\mathcal{C}_{\mathrm{leaf}}|$ & \texttt{atomic} & \texttt{episode} & \texttt{storyline} & \texttt{none} & Audited Provenance / IAA \\
\midrule
X-Pilot & X/Twitter (Dev) & 2 & 6 & 36 & 200$^{a}$ & 35 & 32 & 91 & 8 & 59 & 200 dual-adjudicated \\
X-Expansion & X/Twitter (Held-Out) & 4 + Hard & 14 & 17$^{b}$ & 497 & 299 & 130 & 239 & 48 & 80 & 200 audited ($\kappa=0.472^{c}$) \\
HN-CP1 & Hacker News (Stage-1 Blind) & 3 + Hard & 9 & 14 & 320 & 150 & 21 & 140 & 35 & 124 & 160 dual ($\kappa=0.742$, $\kappa_{\text{Ep/At}}=0.943$) \\
HN-Reserve & Hacker News (Stage-2 Blind)$^{d}$ & 3 + Hard & 9 & 14 & 160 & 72 & 12 & 70 & 6 & 72 & 80 dual ($\kappa=0.852$, $\kappa_{\text{Ep/At}}=0.723$) \\
\midrule
\textbf{Total} & \textbf{2 Platforms} & \textbf{9 + 2 Hard} & \textbf{29} & \textbf{67} & \textbf{1,177} & \textbf{556} & \textbf{195} & \textbf{540} & \textbf{97} & \textbf{335} & \textbf{640 dual-audited records} \\
\bottomrule
\end{tabular}%
}
\par\vspace{0.25em}\raggedright\scriptsize $^{a}$Pilot includes 10 sparse \texttt{multi\_event} diagnostic items (omitted from the 4 core columns). $^{b}$Expansion Event Card \texttt{v3.2} defines 16 core active Atomics plus 1 split sibling leaf invoked by candidate generation (17 predicted leaves across 12 Episodes; 8/12 single-child). $^{c}$Post-freeze independent expert v3 reannotation on the fixed 200-post subset confirmed linkable agreement 0.915 ($\kappa=0.672$) while quantifying genuine Episode-vs-Atomic boundary ambiguity. $^{d}$All 12 Stage-2 prediction artifacts (including frozen \texttt{AdaDepth v1.0} and all IR/NLI baselines) were locked prior to annotation and verified for file consistency before evaluation.
\end{table}

\subsection{Platform 1: X/Twitter Pilot-200 and Held-Out Expansion-497}
The X/Twitter corpus comprises noisy, informal social-media posts collected via event-oriented queries (retaining text, timestamp, media indicator, and deduplication lineage while stripping user profiles):
\begin{itemize}[leftmargin=*,nosep]
  \item \textbf{X-Pilot-200 (Development):} Sampled from 1,728 unique posts across two Argentina World Cup Storylines (6 Episodes, 36 Atomic Events; 100\% multi-child Episodes). Two reviewers independently annotated all 200 items followed by structured adjudication (32 \texttt{atomic\_exact}, 91 \texttt{episode\_only}, 8 \texttt{storyline\_only}, 59 \texttt{none}, 10 \texttt{multi\_event}).
  \item \textbf{X-Expansion-497 (Held-Out):} Drawn from a 4,704-post frame spanning four held-out Storylines (the 2024 CrowdStrike outage: 115, Apple Vision Pro rollout: 112, SpaceX Starship flight tests: 95, and the EU AI Act: 75) plus a mutually exclusive 100-post \nolinkurl{HARD_POOL} (63 of 80 \texttt{none} records; a separate 109-item hard-source flag includes 9 hard-source items inside named pools). Before any Gold label existed, six systems generated frozen predictions over a locked 650-item queue; 497/500 Checkpoint-1 items were eligible. A pre-fixed 200-post audited subset tracks consensus (50) and expert-adjudicated (150) provenance alongside 297 single-reviewed items. Post-freeze structural inspection revealed a strong topology skew in Event Card \texttt{v3.2}: 8 of 12 candidate-active Episodes are single-child ($n_{\mathrm{sib}}=0$), accounting for 235/299 (78.6\%) of leaf proposals and 107/125 (85.6\%) of rescuable over-descends (e.g., \texttt{EP31\_02} airline disruption yields 0/41 Atomic precision).
\end{itemize}

\subsection{Platform 2: Preregistered Hacker News Stories Blind Benchmark (CP1-320)}
To evaluate zero-parameter transfer across platform, genre, time period, and balanced card topology without retrospective leakage, we built a homogeneous \textbf{Hacker News (HN) Stories} benchmark under frozen \textbf{Protocol v2.1}.
Because preliminary probing showed standalone HN comments presuppose thread titles (yielding only 2.5\% \texttt{atomic\_exact}), we collected strictly homogeneous HN \texttt{tags=story} submissions via the official Algolia HN API across three 2024--2025 Storylines grounded in fixed Wikipedia revisions (\texttt{SL25\_DEEPSEEK\_R1}, \texttt{SL25\_TIKTOK\_US\_BAN}, and \texttt{SL24\_STARLINER\_CFT}) plus a topical \texttt{HARD\_POOL} (2,418 unique headlines after deduplicating 2,849 raw records; median 10 words across all four pools).
We authored 9 Episodes (balanced: 4 single-child, 5 multi-child) and 14 active Atomic Events (3 \texttt{low}, 4 \texttt{medium}, 7 \texttt{high} pre-Gold Granularity Gap), each paired with an explicit $\Delta(e,a)$ assertion and disqualifying parent-framing rules.

From a stratified 480-item locked queue (Checkpoint-1: 320 items = 85/85/85/65 across pools; Reserve: 160 items = 40/40/40/40), a frozen one-stage \texttt{Qwen2.5-7B-Instruct} generator emitted 222/480 leaf proposals (0 schema violations; 150/320 in CP1 achieving 100\% recall on all 21 Gold \texttt{atomic\_exact} headlines).
All signal scores were frozen and verified for file consistency \emph{before} Gold annotation began.
CP1-320 underwent dual review on a pre-fixed 160-item overlap subset (42/42/42/34 across pools), achieving 1.000 eligibility agreement, 0.831 four-class agreement ($\kappa=0.742$), 0.984 Episode-vs-Atomic agreement ($\kappa=0.943$, $n=62$), and 0.954 verbatim-span token-F1 ($n=10$).
All 30 disagreements were resolved through structured adjudication (25 Reviewer~A, 5 Reviewer~B), yielding frozen \texttt{HN-CP1-320} and the standalone fully dual-audited \texttt{HN-Dual-160} sensitivity subset ($|\mathcal{C}_{\mathrm{leaf}}|=72$).

\subsection{Stage-2 Prospective Blind Validation: Frozen Reserve-160}\label{sec:reserve160}
Because \texttt{AdaDepth v1.0} was finalized after observing the Granularity-Gap regime shift between X-Expansion and HN-CP1-320, prospective validation requires evaluating the frozen algorithm on newly annotated items whose Gold labels did not exist at design time.
We therefore executed a two-step freeze-then-annotate protocol on the held-out 160 HN Stories items (\texttt{HNS\_0321}--\texttt{HNS\_0480}, stratified 40/40/40/40 across all four pools, containing 72 frozen leaf proposals):
\begin{enumerate}[leftmargin=*,nosep]
  \item \textbf{Pre-Annotation Prediction Freeze and Consistency Check:} Using strictly the 72 unlabeled Reserve leaf candidates (with zero access to CP1 statistics or Gold labels), we ran frozen \texttt{AdaDepth v1.0} (\texttt{Qwen2.5-7B} and \texttt{Phi-4-mini}) and all four modern GPU IR/NLI baselines, recording all 12 prediction artifacts in a pre-annotation consistency registry (\Cref{tab:app_freeze}). On the unlabeled 72 Reserve candidates, \texttt{AdaDepth}'s null-anchored prevalence anchor automatically estimated $\hat{\pi}_L = 0.139$ (10/72 KEEP, 62/72 BACKOFF) for \texttt{Qwen2.5-7B} and $\hat{\pi}_L = 0.097$ (7/72 KEEP, 65/72 BACKOFF) for \texttt{Phi-4-mini}, closely bracketing the subsequently revealed true \texttt{KeepOK} rate of $8/72 = 11.1\%$.
  \item \textbf{Post-Freeze Dual-Review Annotation and Adjudication:} Only after all 12 prediction files were locked did two independent annotators complete \texttt{Reserve-160} under frozen Protocol~v2.1 (Annotator~A: all 160 items; Annotator~B: stratified 80-item overlap subset, 20/20/20/20 across pools, seed \texttt{20261129}; 0 schema/verbatim-span errors on both workbooks). Independent dual agreement on \texttt{Reserve-Dual-80} reached \textbf{1.000 eligibility}, \textbf{0.913 four-class agreement ($\kappa=0.852$)}, \textbf{0.900 exact legal-path agreement (72/80)}, \textbf{0.919 binary Episode-vs-Atomic agreement ($\kappa=0.723$, $n=37$)}, and \textbf{0.980 verbatim-span token-F1 ($n=5$)}. All 8 disagreements (concentrated at the Episode-vs-Atomic boundary on evaluative commentary headlines) were resolved through structured adjudication (3 following Reviewer~A, 5 following Reviewer~B), producing frozen \nolinkurl{hn_stories_reserve_gold_160_v1.0.csv} (72 dual consensus, 8 expert adjudicated, 80 single reviewed; 12 \texttt{atomic\_exact}, 70 \texttt{episode\_only}, 6 \texttt{storyline\_only}, 72 \texttt{none}) and the standalone dual-audited sensitivity subset \texttt{Reserve-Dual-80}. Automated integrity checks confirmed exact file consistency across all 12 pre-annotation prediction files before computing the prospective metrics in \S\ref{sec:stage2_res}. Across the complete 1,177-record benchmark, 640 records (54.4\%) underwent independent dual review (100\% Pilot, 40.2\% X-Exp, 50\% HN-CP1, and 50\% HN-Reserve). Note that \texttt{Reserve-160} shares the same three-Storyline HN ontology as \texttt{HN-CP1-320}, so it tests prospective transfer to unseen items and post-freeze Gold labels under a fixed event ontology rather than a fourth unseen Storyline ontology.
\end{enumerate}

\section{Methodology: The \texorpdfstring{\texttt{EviLink}}{EviLink} Framework and \texorpdfstring{\texttt{AdaDepth}}{AdaDepth} Calibration}\label{sec:method}
\begin{figure}[t]
\centering
\includegraphics[width=\textwidth]{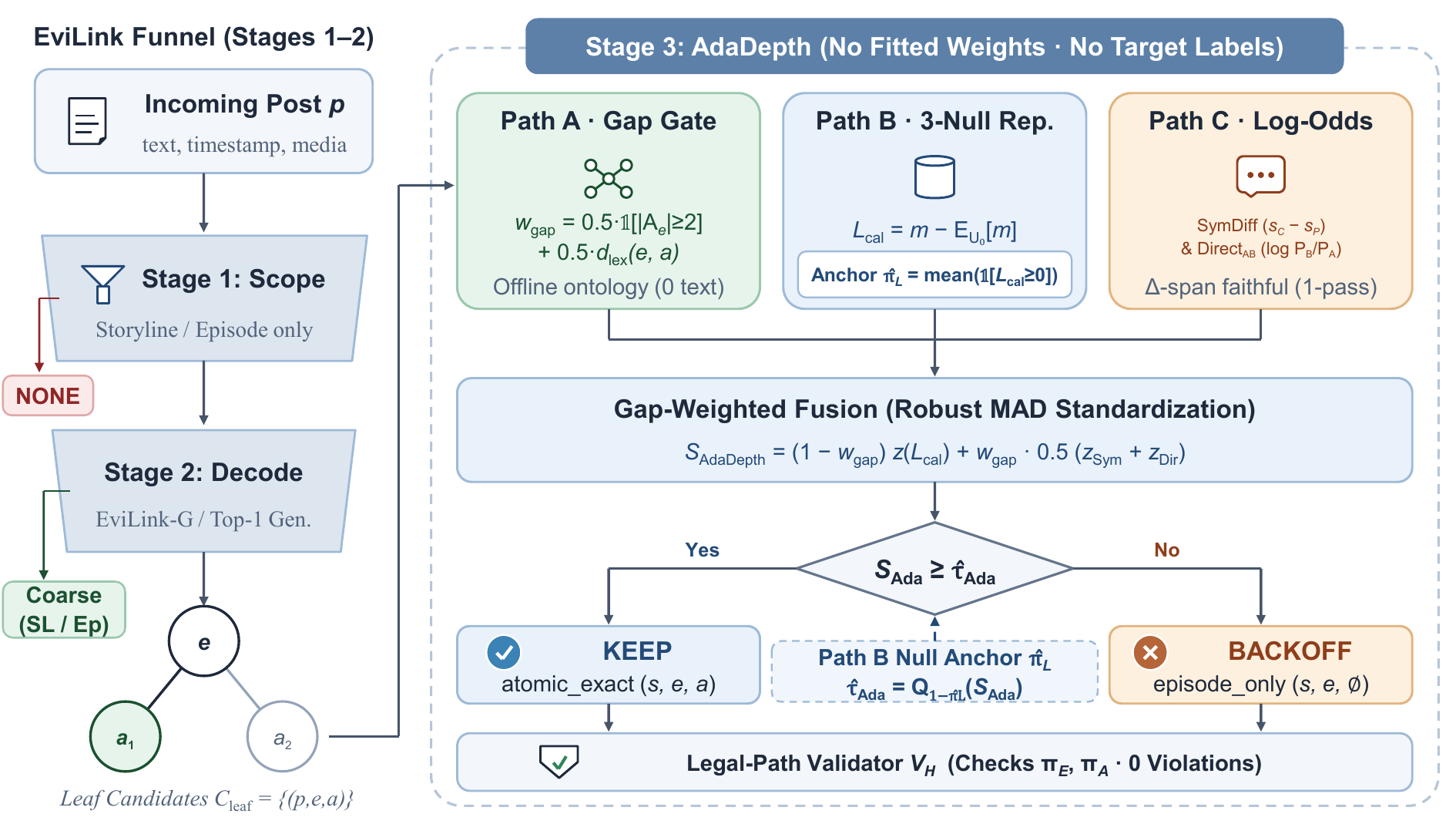}
\caption{The unified \textbf{\texttt{EviLink}} architecture separating the upstream open-set funnel (Stages~1--2) from \textbf{Stage~3 \texttt{AdaDepth}} (Granularity-Gap-Aware, Null-Anchored Calibrator). Path~A computes an offline card-intrinsic structural gap weight $w_{\mathrm{gap}}(e,a)$ with zero text or target labels; Path~B centers final-layer representation margins against three content-free nulls to produce both $z(L_{\mathrm{cal}})$ and the unlabeled prevalence anchor $\hat{\pi}_L$; Path~C extracts $\Delta(e,a)$-faithful prompt log-odds (\texttt{SymDiff}, \texttt{Direct\_AB}). Gap-weighted fusion and quantile thresholding yield legal KEEP/BACKOFF outputs verified by $V_{\mathcal{H}}$.}
\label{fig:method}
\Description{System architecture diagram showing the EviLink Upstream Funnel (Stage 1 Scope Gate and Stage 2 Path Decoder) feeding leaf candidates into Stage 3 AdaDepth, which combines Path A structural gap gate, Path B 3-null representation centering and prevalence anchor, and Path C prompt log-odds through a quantile gate into KEEP or BACKOFF decisions verified by V_H.}
\end{figure}

\Cref{fig:method} depicts the unified three-stage \textbf{\texttt{EviLink}} architecture, decoupling open-set membership filtering (Stage~1) and legal path decoding (Stage~2) from zero-parameter stopping-depth calibration (Stage~3, \texttt{AdaDepth}).

\subsection{Stages 1--2: Open-set scope gate, hierarchy path decoders, and validator}
To prevent fine-grained leaf keywords from causing false attachment on out-of-inventory posts, \texttt{EviLink} separates scope filtering from leaf decoding:
\begin{itemize}[leftmargin=*,nosep]
  \item \textbf{Stage 1 (Open-Set Scope Gate):} Given post $p$ and \emph{only} Storyline and Episode Cards $(\mathcal{S},\mathcal{E})$ (withholding Atomic Cards), the gate classifies $g(p)\in\{\mathrm{IN\_SCOPE}, \mathrm{NONE}\}$. Rejected posts exit immediately as $(s=e=a=\mathrm{NONE}, \texttt{none})$.
  \item \textbf{Stage 2 (Hierarchy Path Decoder):} For in-scope posts, \textbf{\texttt{EviLink-G}} feeds the full frozen registry $\mathcal{H}$ for global path reasoning, while \textbf{\texttt{EviLink-S}} first routes to $(s,e)$ and then scores local Atomic candidates $a\in\mathcal{A}_e$. On HN Stories, a frozen open-weight one-stage generator (\texttt{Qwen2.5-7B-Instruct}) directly emits structured tuples over neutral IDs. Coarse predictions (\texttt{storyline\_only}, \texttt{episode\_only}) pass straight to validator $V_{\mathcal{H}}$; leaf proposals enter $\mathcal{C}_{\mathrm{leaf}}$.
\end{itemize}

\subsection{Stage 3: Elementary zero-parameter stopping-depth signals}
Over any leaf proposal $(p,e,a)\in\mathcal{C}_{\mathrm{leaf}}$ scored by an open-weight instruction-tuned LLM (we evaluate \texttt{Qwen2.5-7B-Instruct} \citep{qwen2024qwen25} and \texttt{Phi-4-mini-instruct} \citep{microsoft2025phi4mini}), we construct two complementary zero-parameter signal pathways:

\textbf{Path B — 3-Null-Centered Representation Margin ($L_{\mathrm{cal}}$).}
Let $\phi(x)\in\mathbb{S}^{d-1}$ be the $L_2$-normalized final-layer mean-pooled hidden representation of text $x$ (layer ablated in \Cref{app:layers}).
The raw Child-minus-Parent cosine margin is
\begin{equation}\label{eq:lraw}
m(p;e,a) = \langle \phi(p), \phi(c_a) \rangle - \langle \phi(p), \phi(c_e) \rangle.
\end{equation}
Because Event Cards vary widely in length, specificity, and formatting boilerplate, $m(p;e,a)$ suffers from static card-pair offset bias independent of post $p$.
Inspired by contextual calibration \citep{zhao2021calibrate}, we center each card pair $(e,a)$ against three content-free null probes $\mathcal{U}_0 = \{\texttt{""}, \texttt{"N/A"}, \texttt{"[No content]"}\}$:
\begin{equation}\label{eq:lcal}
L_{\mathrm{cal}}(p;e,a) = m(p;e,a) - b_0(e,a), \qquad b_0(e,a) = \frac{1}{|\mathcal{U}_0|}\sum_{u\in\mathcal{U}_0} m(u;e,a).
\end{equation}
In practice, card embeddings $\phi(c_e), \phi(c_a)$ and null offsets $b_0(e,a)$ are computed once \emph{offline} per registry, requiring only a single encoder forward pass per incoming post at inference time.

\textbf{Path C — Incremental-Evidence Prompt Log-Odds (\texttt{SymDiff}, \texttt{Direct\_AB}).}
Representation mean-pooling captures broad topical alignment but dilutes single-clause incremental assertions.
We therefore extract two zero-shot next-token log-odds signals directly from the LLM's first-token logit head (1 forward pass each, no autoregressive decoding):
\begin{itemize}[leftmargin=*,nosep]
  \item \textbf{Decoupled Symmetric Difference (\texttt{SymDiff}):} Two independent binary Yes/No queries ask (i) whether Atomic claim $c_a$ is the post's central main proposition, giving $s_C(p,a) = \log P(\text{Yes}) - \log P(\text{No})$, and (ii) whether $p$ is broader reaction/impact commentary better suited to Episode scope $c_e$, giving $s_P(p,e) = \log P(\text{Yes}) - \log P(\text{No})$. We form $\mathrm{SymDiff}(p;e,a) = s_C(p,a) - s_P(p,e)$ (alongside standalone Child $s_C$ and Parent-Scope Rejection $\mathrm{Parent\_inv} = -s_P$).
  \item \textbf{Contrastive A/B Log-Odds (\texttt{Direct\_AB}):} A single contrastive prompt presents post $p$, Parent $c_e$ (Option~A), and Child $c_a$ (Option~B), scoring $\mathrm{Direct}_{AB}(p;e,a) = \log P(\text{B}) - \log P(\text{A})$.
\end{itemize}

\subsection{\texorpdfstring{\texttt{AdaDepth}}{AdaDepth}: Granularity-Gap-Aware, Null-Anchored Adaptive Calibration}
As our empirical analysis (\S\ref{sec:mech}) reveals, neither Path~B nor Path~C dominates universally: $L_{\mathrm{cal}}$ excels when parent and child cards have narrow \emph{low} Granularity Gap on noisy discourse, whereas prompt log-odds (\texttt{SymDiff}, \texttt{Direct\_AB}) dominate when parent and child cards have crisp \emph{high} Granularity Gap. However, raw prompt log-odds suffer severe cross-model threshold drift at $\tau=0$ (e.g., \texttt{Phi-4-mini}'s \texttt{Direct\_AB} keeps 96/97 HN over-descends at $\tau=0$ despite 0.877~AUC).
\textbf{\texttt{AdaDepth}} resolves both obstacles simultaneously with \emph{zero fitted parameters} and \emph{zero target labels}:

\textbf{1. Card-Intrinsic Structural Granularity-Gap Gate (Path~A).}
For every candidate pair $(e,a)$, we compute an offline, post-independent structural gap weight $w_{\mathrm{gap}}(e,a)\in[0,1]$ solely from ontology topology and card vocabulary:
\begin{equation}\label{eq:wgap}
w_{\mathrm{gap}}(e,a) = 0.5\,\mathbb{I}\bigl[|\mathcal{A}_e| \ge 2\bigr] + 0.5\,d_{\mathrm{lex}}(e,a), \qquad d_{\mathrm{lex}}(e,a) = \frac{|W(c_a) \setminus W(c_e)|}{\max(|W(c_a)|, 1)},
\end{equation}
where $W(\cdot)$ is the lowercased alphanumeric content-word set (length $\ge 3$, stopwords removed).
Single-child episodes whose Atomic card largely paraphrases the Episode yield low $w_{\mathrm{gap}}$, prioritizing noise-resistant representation centering $L_{\mathrm{cal}}$; multi-child episodes with distinct novel Atomic vocabulary yield $w_{\mathrm{gap}}\to 1$, shifting weight cleanly toward $\Delta(e,a)$-faithful prompt log-odds.

\textbf{2. Robust MAD Standardization and Gap-Weighted Fusion.}
Because representation margins ($\sim 10^{-2}$) and prompt log-odds ($\sim 10^1$) live on incommensurate scales, we standardize each signal $x\in\{L_{\mathrm{cal}}, \mathrm{SymDiff}, \mathrm{Direct}_{AB}\}$ across the unlabeled leaf-candidate batch $\mathcal{C}_{\mathrm{leaf}}$ using outlier-resistant Median Absolute Deviation (MAD):
\begin{equation}\label{eq:mad}
z(x_i) = \frac{x_i - \mathrm{median}(x)}{\hat{\sigma}(x)}, \qquad \mathrm{MAD}(x) = \mathrm{median}\bigl(|x - \mathrm{median}(x)|\bigr),
\end{equation}
where $\hat{\sigma}(x) = 1.4826\,\mathrm{MAD}(x)$ if $\mathrm{MAD}(x) > 0$ and $\hat{\sigma}(x) = \mathrm{std}(x)$ otherwise (setting $z(x_i)=0$ for constant vectors with $\hat{\sigma}(x)=0$, using $10^{-8}$ numerical tolerance in implementation).
Setting $z_{\mathrm{prm}}(i) = 0.5\,\bigl(z(\mathrm{SymDiff}_i) + z(\mathrm{Direct}_{AB,i})\bigr)$, the fused \texttt{AdaDepth} sufficiency score is
\begin{equation}\label{eq:adadepth_score}
S_{\mathrm{AdaDepth}}(i) = \bigl(1 - w_{\mathrm{gap}}(\hat{e}_i,\hat{a}_i)\bigr)\,z(L_{\mathrm{cal},i}) + w_{\mathrm{gap}}(\hat{e}_i,\hat{a}_i)\,z_{\mathrm{prm}}(i).
\end{equation}

\textbf{3. Null-Anchored Prevalence Operating Point.}
Standardized $z$-scores have zero median, so thresholding $S_{\mathrm{AdaDepth}}\ge 0$ would blindly keep 50\% of leaf candidates regardless of whether true Atomic prevalence is 10\% (HN Stories) or 66\% (X-Pilot).
Instead, we exploit a key empirical property of 3-null representation centering: the unlabeled non-negative rate of $L_{\mathrm{cal}}$,
\begin{equation}\label{eq:pil}
\hat{\pi}_L = \frac{1}{|\mathcal{C}_{\mathrm{leaf}}|}\sum_{i\in\mathcal{C}_{\mathrm{leaf}}} \mathbb{I}\bigl[L_{\mathrm{cal}}(p_i;\hat{e}_i,\hat{a}_i) \ge 0\bigr],
\end{equation}
tracks the true domain base rate across a $6\times$ shift with zero labels (e.g., on \texttt{Qwen2.5-7B}: $\hat{\pi}_L = 0.153$ on HN-320 [true 10.7\%], $0.458$ on X-Exp [true 35.8\%], and $0.943$ on X-Pilot [true 65.7\%]).
\texttt{AdaDepth} therefore transfers this calibrated prevalence budget directly to the higher-ranking fused score via the $(1-\hat{\pi}_L)$ empirical quantile:
\begin{equation}\label{eq:tau_ada}
\hat{\tau}_{\mathrm{Ada}} = \mathrm{Quantile}_{1-\hat{\pi}_L}\bigl(\{S_{\mathrm{AdaDepth}}(i)\}_{i\in\mathcal{C}_{\mathrm{leaf}}}\bigr),
\end{equation}
retaining top-$\hat{\pi}_L$ candidates as \texttt{atomic\_exact} and backing off the rest to \texttt{episode\_only}.
Standardization (\Cref{eq:mad}), prevalence estimation (\Cref{eq:pil}), and quantile thresholding (\Cref{eq:tau_ada}) compute batch-level statistics over unlabeled leaf candidates $\mathcal{C}_{\mathrm{leaf}}$ (or a rolling-window buffer in online streams, falling back to standard deviation if $\mathrm{MAD}=0$); Appendix~\Cref{tab:app_robust}(c) confirms stability down to $B=16$ ($\mathrm{std}\le 0.007$, stabilizing for $B\ge 32$).
\Cref{alg:adadepth} summarizes the complete zero-parameter procedure.

\begin{algorithm}[t]
\small
\caption{\texttt{AdaDepth}: Zero-Parameter Granularity-Gap-Aware Stopping-Depth Calibration}
\label{alg:adadepth}
\KwIn{Frozen ontology $\mathcal{H}$, unlabeled leaf candidate batch $\mathcal{C}_{\mathrm{leaf}} = \{(p_i,\hat{s}_i,\hat{e}_i,\hat{a}_i)\}_{i=1}^M$, open-weight LLM $\mathcal{M}$, null probes $\mathcal{U}_0=\{\texttt{""},\texttt{"N/A"},\texttt{"[No content]"}\}$.}
\KwOut{Validated hierarchical linking predictions $\{\hat{y}_i\}_{i=1}^M$.}
\BlankLine
\tcp{Offline per-registry pass (computed once for $\mathcal{H}$)}
\ForEach{active parent--child pair $(e,a)$ with $\pi_A(a)=e$}{
  Compute $w_{\mathrm{gap}}(e,a) \leftarrow 0.5\,\mathbb{I}[|\mathcal{A}_e|\ge 2] + 0.5\,|W(c_a)\setminus W(c_e)|/\max(|W(c_a)|,1)$\;
  Cache card embeddings $\phi(c_e),\phi(c_a)$ and 3-null offset $b_0(e,a) \leftarrow \frac{1}{3}\sum_{u\in\mathcal{U}_0} \langle\phi(u),\phi(c_a)-\phi(c_e)\rangle$\;
}
\tcp{Online inference pass over unlabeled leaf candidates $\mathcal{C}_{\mathrm{leaf}}$}
\For{$i = 1, \dots, M$}{
  $L_{\mathrm{cal},i} \leftarrow \langle\phi(p_i),\phi(c_{\hat{a}_i})-\phi(c_{\hat{e}_i})\rangle - b_0(\hat{e}_i,\hat{a}_i)$\;
  Extract prompt log-odds $\mathrm{SymDiff}_i \leftarrow s_C(p_i,\hat{a}_i) - s_P(p_i,\hat{e}_i)$ and $\mathrm{Direct}_{AB,i} \leftarrow \log P(\text{B}|p_i) - \log P(\text{A}|p_i)$\;
}
Standardize $z(L_{\mathrm{cal}}), z(\mathrm{SymDiff}), z(\mathrm{Direct}_{AB})$ across $i=1,\dots,M$ via robust MAD (\Cref{eq:mad})\;
$\hat{\pi}_L \leftarrow \frac{1}{M}\sum_{i=1}^M \mathbb{I}[L_{\mathrm{cal},i} \ge 0]$\;
\For{$i = 1, \dots, M$}{
  $z_{\mathrm{prm},i} \leftarrow 0.5\,\bigl(z(\mathrm{SymDiff}_i)+z(\mathrm{Direct}_{AB,i})\bigr)$\;
  $S_i \leftarrow \bigl(1 - w_{\mathrm{gap}}(\hat{e}_i,\hat{a}_i)\bigr)\,z(L_{\mathrm{cal},i}) + w_{\mathrm{gap}}(\hat{e}_i,\hat{a}_i)\,z_{\mathrm{prm},i}$\;
}
$\hat{\tau}_{\mathrm{Ada}} \leftarrow \mathrm{Quantile}_{1-\hat{\pi}_L}(\{S_i\}_{i=1}^M)$\;
\For{$i = 1, \dots, M$}{
  \lIf{$S_i \ge \hat{\tau}_{\mathrm{Ada}}$}{$\hat{y}_i \leftarrow V_{\mathcal{H}}(\hat{s}_i,\hat{e}_i,\hat{a}_i,\texttt{atomic\_exact})$}
  \lElse{$\hat{y}_i \leftarrow V_{\mathcal{H}}(\hat{s}_i,\hat{e}_i,\varnothing,\texttt{episode\_only})$}
}
\Return{$\{\hat{y}_i\}_{i=1}^M$}\;
\end{algorithm}

\subsection{Theoretical properties of \texorpdfstring{\texttt{AdaDepth}}{AdaDepth}}
Two concise structural properties explain why \texttt{AdaDepth} remains stable across unseen domains and architectures without tuning:

\begin{proposition}[Exact Within-Pair Invariance and Additive Card-Bias Removal]\label{prop:inpair}
For any fixed card pair $(e,a)$ and any two posts $p_1, p_2$ assigned to $(e,a)$, 3-null centering satisfies $L_{\mathrm{cal}}(p_1;e,a) - L_{\mathrm{cal}}(p_2;e,a) = m(p_1;e,a) - m(p_2;e,a)$, hence $\mathrm{InPairAUC}(L_{\mathrm{cal}}) \equiv \mathrm{InPairAUC}(L_{\mathrm{raw}})$ identically.
Moreover, if the raw hidden-state margin decomposes as $m(p;e,a) = \mu_{\mathrm{ev}}(p;e,a) + \beta_{e,a}$ where $\beta_{e,a}$ is any post-independent card-pair length/boilerplate offset and null probes satisfy $\mathbb{E}_{u\in\mathcal{U}_0}[\mu_{\mathrm{ev}}(u;e,a)]=0$, then $L_{\mathrm{cal}}(p;e,a) = \mu_{\mathrm{ev}}(p;e,a)$ recovers the unbiased evidential margin across heterogeneous card pairs.
\end{proposition}
\begin{proof}
Since $b_0(e,a)$ depends solely on $(e,a)$ and not on $p$, subtracting $b_0(e,a)$ applies a constant shift within every pair $(e,a)$, preserving all strict pairwise orderings $\mathbb{I}[L_{\mathrm{cal}}(p_1)>L_{\mathrm{cal}}(p_2)] = \mathbb{I}[m(p_1)>m(p_2)]$ and therefore $\mathrm{InPairAUC}$ exactly.
Under $m(p;e,a) = \mu_{\mathrm{ev}}(p;e,a) + \beta_{e,a}$, averaging over $\mathcal{U}_0$ yields $b_0(e,a) = \beta_{e,a} + \frac{1}{|\mathcal{U}_0|}\sum_{u\in\mathcal{U}_0}\mu_{\mathrm{ev}}(u;e,a) = \beta_{e,a}$, canceling $\beta_{e,a}$ identically.
\end{proof}

\begin{proposition}[Decoupled Prevalence-Anchored Quantile Control and Exact HEM Decomposition]\label{prop:quantile}
Let $S_{\mathrm{AdaDepth}}$ be the fused score whose threshold is set by $\hat{\tau}_{\mathrm{Ada}} = \mathrm{Quantile}_{1-\hat{\pi}_L}(S_{\mathrm{AdaDepth}})$, and let $\tilde{x}_i = \alpha x_i + \gamma$ be any strictly positive affine drift ($\alpha>0,\gamma\in\mathbb{R}$) affecting a non-constant prompt log-odds component $x\in\{\mathrm{SymDiff}, \mathrm{Direct}_{AB}\}$ ($\hat{\sigma}(x) > 0$) due to instruction-tuning bias or domain shift.
For any decision rule retaining a leaf subset $\mathcal{K}\subseteq\mathcal{C}_{\mathrm{leaf}}$ of size $K = |\mathcal{K}|$, let $k_1 = \sum_{i\in\mathcal{K}}\mathbb{I}[\mathrm{KeepOK}_i=1]$, $k_2 = \sum_{i\in\mathcal{K}}\mathbb{I}[\mathrm{BackoffOK}_i=1]$, and $k_3 = K - k_1 - k_2$ denote the kept counts of correct leaves, rescuable over-descends, and upstream errors.
Then:
(i)~the KEEP/BACKOFF decision vector $\mathrm{Dec}_{\hat{\tau}_{\mathrm{Ada}}}$ is invariant to $(\alpha,\gamma)$;
(ii)~end-to-end exact match obeys the exact identity $\mathrm{HEM}(\mathcal{K}) - \mathrm{HEM}_{\mathrm{All\text{-}BO}} = (k_1 - k_2)/N$; consequently, at identical keep budget $K$ (specificity $\bar{D}=2+K/|\mathcal{C}_{\mathrm{leaf}}|$), higher \texttt{KeepOK} retention ($k_1$) strictly guarantees higher $\mathrm{HEM}$ on the decisive subset ($k_3=0$, where $k_2 = K - k_1$), and more generally in open-set streams whenever kept \texttt{BackoffOK} count does not increase faster than \texttt{KeepOK} ($\Delta k_1 > \Delta k_2$).
\end{proposition}
\begin{proof}
(i)~Under any positive affine shift $\tilde{x}_i = \alpha x_i + \gamma$ ($\alpha>0$) on a non-constant vector ($\hat{\sigma}(x)>0$), $\mathrm{median}(\tilde{x}) = \alpha\,\mathrm{median}(x)+\gamma$, $\mathrm{MAD}(\tilde{x}) = \alpha\,\mathrm{MAD}(x)$, and $\mathrm{std}(\tilde{x}) = \alpha\,\mathrm{std}(x)$, hence $\hat{\sigma}(\tilde{x}) = \alpha\,\hat{\sigma}(x)$ and $z(\tilde{x}_i) = \frac{\alpha(x_i-\mathrm{median}(x))}{\alpha\,\hat{\sigma}(x)} = z(x_i)$ identically. Thus $S_{\mathrm{AdaDepth}}$ is invariant, and because $\hat{\pi}_L$ depends solely on $L_{\mathrm{cal}}$, $\mathrm{Dec}_{\hat{\tau}_{\mathrm{Ada}}}$ is invariant to $(\alpha,\gamma)$.
(ii)~Relative to unconditional All-Backoff (which scores exact match on all \texttt{BackoffOK} items and zero on \texttt{KeepOK} and upstream-error items), retaining $\mathcal{K}$ gains $+1$ exact match on each of the $k_1$ kept \texttt{KeepOK} items, loses $-1$ exact match on each of the $k_2$ kept \texttt{BackoffOK} items, and incurs $0$ change on the $k_3$ kept upstream errors, yielding $\mathrm{HEM}(\mathcal{K}) - \mathrm{HEM}_{\mathrm{All\text{-}BO}} = (k_1 - k_2)/N$.
When $k_3=0$ (no upstream errors in $\mathcal{C}_{\mathrm{leaf}}$), $k_2 = K - k_1$ gives $\mathrm{HEM}(\mathcal{K}) - \mathrm{HEM}_{\mathrm{All\text{-}BO}} = (2k_1 - K)/N$, which is strictly increasing in $k_1$ at fixed $K$.
In open-set streams with $k_3 > 0$, substituting a kept upstream error ($k_3$, cost $0$) with a \texttt{BackoffOK} item ($k_2$, cost $-1$) can offset a smaller increase in $k_1$; thus net exact-match improvement requires $\Delta(k_1 - k_2) > 0$.
\end{proof}

\section{Experimental Setup and Baselines}\label{sec:setup}
\textbf{End-to-end upstream linkers (Platform~1).}
On X/Twitter Pilot-200 and held-out Expansion-497, we benchmark flat lexical \texttt{BM25}, dense \texttt{BGE-small}, \texttt{MiniLM-L6} cross-encoder, a Pilot-threshold-optimized Hierarchical Cross-Encoder, \texttt{Full Cards LLM}, \texttt{EviLink-S}, \texttt{EviLink-G}, and a Pilot-trained meta-router \texttt{EviLink-R}, all using the frozen hosted LLM snapshot (\texttt{deepseek-v4-flash}).
On HN Stories (Platform~2), Stage~2 uses the frozen open-weight \texttt{Qwen2.5-7B-Instruct} one-stage candidate generator.

\textbf{Stopping-depth calibration baselines (15+ comparators across 6 families).}
Over identical frozen leaf candidate sets $\mathcal{C}_{\mathrm{leaf}}$ across all four Gold evaluation sets (\texttt{X-Expansion-497}, \texttt{X-Pilot-200}, \texttt{HN-CP1-320}, \texttt{HN-Dual-160}), we evaluate:
\begin{enumerate}[leftmargin=*,nosep]
  \item \textbf{Lexical IR:} Zero-parameter \texttt{Okapi BM25} margin $\mathrm{BM25}(p,c_a) - \mathrm{BM25}(p,c_e)$ \citep{robertson2009}.
  \item \textbf{Modern Dense Bi-Encoder IR:} \texttt{BAAI/bge-large-en-v1.5} \citep{xiao2024cpack} cosine margin $\langle e(p), e(c_a)\rangle - \langle e(p), e(c_e)\rangle$, evaluated both raw and with 3-null centering.
  \item \textbf{Neural Cross-Encoder Reranker:} \nolinkurl{cross-encoder/ms-marco-MiniLM-L-6-v2} \citep{wang2020minilm} relevance margin $r(p,c_a) - r(p,c_e)$ (raw and 3-null centered).
  \item \textbf{Modern NLI Entailment Classifiers:} \nolinkurl{MoritzLaurer/DeBERTa-v3-large-mnli-fever-anli-ling-wanli} \citep{he2023debertav3,laurer2024less} and \nolinkurl{facebook/bart-large-mnli} \citep{lewis2020bart}, scoring Child-minus-Parent entailment log-odds margin (raw and 3-null centered).
  \item \textbf{Selective Prediction \& Hierarchical Hedging:} Maximum Softmax Probability (\textbf{MSP}, keeping when $\max(P_A,P_B)\ge 0.75$), \textbf{Negative Entropy} (keeping when binary entropy $H\le 0.5$~bits) \citep{geifman2017selective}, and hierarchical accuracy--specificity hedging (\textbf{DARTS}; \citealp{deng2012hedging}) using A/B option posteriors $P(\text{leaf}) = \sigma(\mathrm{Direct}_{AB})$. Following \citet{deng2012hedging}, DARTS maximizes expected information gain $P(\text{leaf})\cdot I_{\mathrm{atomic}}$ vs.\ $1.0\cdot I_{\mathrm{episode}}$, keeping the leaf iff $P(\text{leaf}) \ge \lambda_{\mathrm{DARTS}} = \log_2|\mathcal{E}|/\log_2|\mathcal{A}|$ ($0.931$ on X-Exp, $0.500$ on X-Pilot, and $0.832$ on HN Stories), with full curve sweep for AURC. Under matched-budget $\hat{\pi}_L$ evaluation, we audit both inclusive quantile thresholding ($S \ge \hat{\tau}$) and strict exact-$K$ uniform tie-breaking to account for discrete logit ties.
  \item \textbf{LLM Diagnostic \& Supervised Comparators:} Free-form evidence self-audit (\texttt{T2}), supervised leave-one-storyline-out (LOSO) regularized logistic regression, individual zero-parameter signals ($L_{\mathrm{cal}}$, \texttt{SymDiff}, \texttt{Direct\_AB}), and a fixed-weight ablation (\textbf{\texttt{Fixed-0.5} + $\hat{\pi}_L$}, setting $w_{\mathrm{gap}}\equiv 0.5$) to isolate the contribution of ontology-intrinsic gap gating in \texttt{AdaDepth}.
\end{enumerate}
All paired comparisons report 10,000-resample paired bootstrap 95\% confidence intervals (seed \texttt{20261128}) and exact two-sided McNemar tests on discordant outcomes \citep{efron1993}. All GPU benchmarks run on a single NVIDIA RTX~PRO~4500 GPU with VRAM and throughput logged directly.

\section{Results I: End-to-End Paradox and Beyond Semantic Relevance}\label{sec:res_part1}

\subsection{End-to-end linking, scope-gate transfer, and the All-Backoff Paradox}
\begin{figure}[t]
\centering
\includegraphics[width=\textwidth]{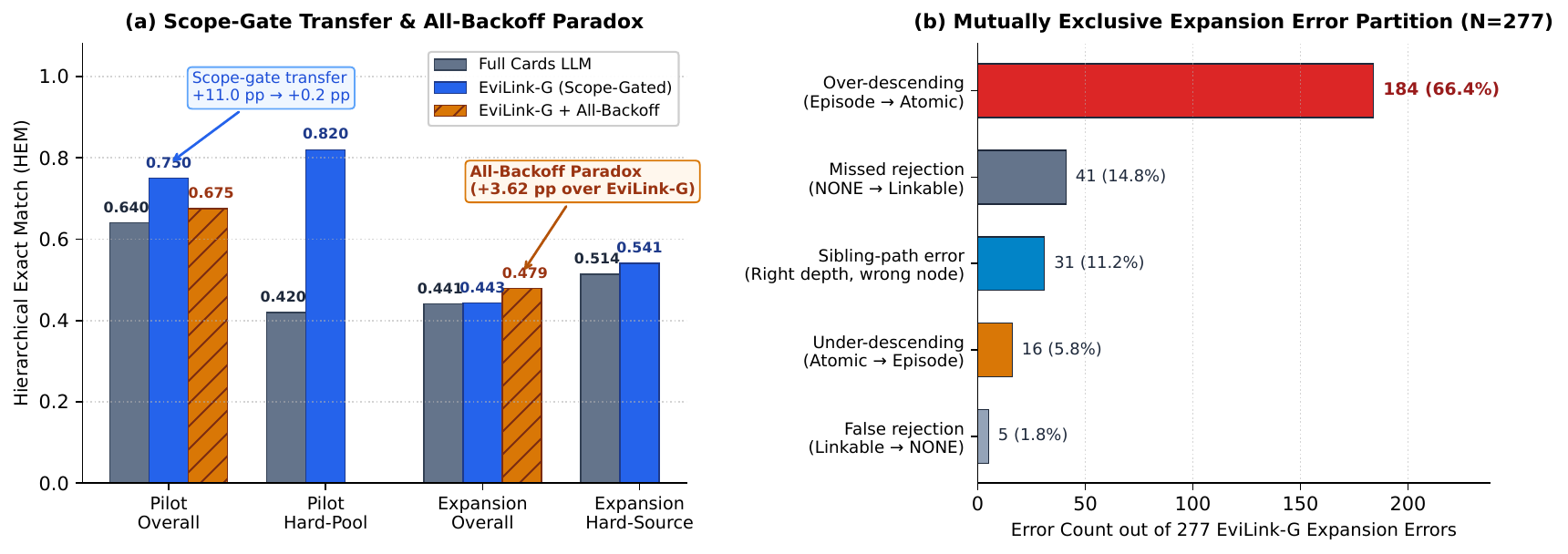}
\caption{Diagnostic origin of the stopping-depth bottleneck on Platform~1 (X/Twitter). \textbf{(a)}~Stage~1 scope gating adds $+11.0$~pp HEM on Pilot-200 ($+40.0$~pp on Hard Pool) but narrows to $+0.2$~pp on held-out Expansion-497, where blanket \emph{All-Backoff} beats \texttt{EviLink-G} by $+3.62$~pp yet harms Pilot by $-7.50$~pp. \textbf{(b)}~Over-descending accounts for 66.4\% (184/277) of all \texttt{EviLink-G} Expansion errors.}
\label{fig:funnel_diag}
\Description{Two-panel diagnostic chart showing Stage 1 scope-gate transfer and the All-Backoff Paradox on the left, and the 277-error partition on X/Twitter Expansion showing 66.4 percent over-descending errors on the right.}
\end{figure}

\Cref{tab:x_end_to_end} and \Cref{fig:funnel_diag} trace end-to-end performance on Platform~1 (X/Twitter).
Three findings stand out.
First, structured LLM path reasoning nearly doubles flat neural retrieval on held-out Expansion-497: \texttt{EviLink-G} reaches 0.443~HEM and \texttt{Full Cards LLM} reaches 0.441~HEM vs.\ 0.248 for the best flat retriever (\texttt{MiniLM-L6} cross-encoder) and 0.161 for \texttt{BM25}.
Second, naive Pilot-tuned thresholding fails under storyline shift: the Pilot-optimized Hierarchical Cross-Encoder rises from 0.275 to 0.330 on Pilot but drops to 0.215 on held-out Expansion, and the Pilot-trained router \texttt{EviLink-R} exhibits negative transfer (0.423~HEM, 95\% CI $[-0.034, -0.002]$ vs.\ Full Cards).
Third, although Stage~1 scope gating yields a $+11.0$~pp jump on Pilot ($+40.0$~pp on Hard Pool) and delivers the top NONE~F1 (0.629) with zero structural violations on Expansion, its net Expansion HEM gain narrows to $+0.2$~pp because \textbf{almost all remaining errors occur inside covered events}.

\begin{table}[t]
\centering
\caption{End-to-end frozen evaluation on Platform~1: X/Twitter Pilot-200 (Dev) and held-out Expansion-497 (plus 200-post audited sensitivity subset). Bold denotes best executable single-system estimate per column.}
\label{tab:x_end_to_end}
\footnotesize
\setlength{\tabcolsep}{3.4pt}
\resizebox{\textwidth}{!}{%
\begin{tabular}{lcccccccccc}
\toprule
& \multicolumn{3}{c}{\textbf{Pilot-200 (Dev)}} & \multicolumn{6}{c}{\textbf{Expansion-497 (Held-Out)}} & \textbf{Audited-200} \\
\cmidrule(lr){2-4}\cmidrule(lr){5-10}\cmidrule(lr){11-11}
Method & HEM & NONE F1 & Hard HEM & HEM & Macro-F1 & NONE F1 & SL Acc. & Ep. Acc. & Hard-Src HEM & HEM \\
\midrule
BM25 & 0.400 & 0.450 & 0.380 & 0.161 & 0.184 & 0.275 & 0.410 & 0.225 & 0.376 & 0.190 \\
BGE-small & 0.250 & 0.400 & 0.438 & 0.235 & 0.199 & 0.371 & 0.391 & 0.260 & 0.578 & 0.285 \\
Cross-encoder & 0.275 & 0.400 & 0.594 & 0.248 & 0.246 & 0.398 & 0.496 & 0.325 & \textbf{0.587} & 0.320 \\
Hier. Cross-enc. ($\tau$-opt.) & 0.330 & 0.392 & 0.380 & 0.215 & 0.185 & 0.347 & 0.357 & 0.224 & 0.367 & -- \\
Full Cards LLM & 0.640 & 0.588 & 0.420 & 0.441 & \textbf{0.430} & 0.615 & \textbf{0.998} & \textbf{0.867} & 0.514 & \textbf{0.440} \\
EviLink-S & 0.620 & \textbf{0.815} & 0.760 & 0.406 & 0.418 & \textbf{0.629} & 0.988 & 0.827 & 0.505 & 0.385 \\
EviLink-G & \textbf{0.750} & \textbf{0.815} & \textbf{0.820} & \textbf{0.443} & 0.424 & \textbf{0.629} & 0.988 & 0.862 & 0.541 & \textbf{0.440} \\
\midrule
EviLink-R (Pilot router)$^{a}$ & -- & -- & -- & 0.423 & -- & -- & -- & -- & -- & -- \\
EviLink-G + All-Backoff & 0.675 & -- & -- & 0.479 & -- & -- & -- & -- & -- & -- \\
Keep-vs-Backoff Oracle & 0.790 & -- & -- & 0.694 & -- & -- & -- & -- & -- & -- \\
\bottomrule
\end{tabular}%
}
\par\vspace{0.25em}\raggedright\scriptsize $^{a}$Frozen negative-transfer diagnostic (95\% paired bootstrap CI vs.\ Full Cards: $[-0.034, -0.002]$). Removing open-set rules on Pilot collapses NONE F1 to 0.000 (0.495 HEM); removing hierarchy constraints produces 123 raw illegal tuples.
\end{table}

Partitioning all 277 \texttt{EviLink-G} Expansion errors (\Cref{fig:funnel_diag}b) exposes the true bottleneck: \textbf{184/277 (66.4\%) are over-descending errors}, dwarfing missed rejections (41, 14.8\%), sibling-path errors (31, 11.2\%), under-descending (16, 5.8\%), and false rejections (5, 1.8\%).
Across the 299 leaf proposals ($\mathrm{KeepOK}=107$, $\mathrm{OD}_{\mathrm{raw}}=138$, $\mathrm{BackoffOK}=125$), blindly reverting every leaf to \texttt{episode\_only} jumps Expansion HEM from 0.4427 to 0.4789 ($+3.62$~pp), and on blind HN Stories ($|\mathcal{C}_{\mathrm{leaf}}|=150$, $\mathrm{KeepOK}=16$, $\mathrm{OD}_{\mathrm{raw}}=97$, $\mathrm{BackoffOK}=58$) jumps HEM from 0.3656 to 0.4969 ($+13.13$~pp). Yet the exact same All-Backoff rule lowers Pilot-200 HEM ($0.7500 \to 0.6750$, $-7.50$~pp).
Leave-one-storyline-out logistic regression on Expansion reaches \textbf{0.3689~AUC} and \textbf{0.4105~HEM}; true leaf-support priors swing from 16\% (CrowdStrike) to 73\% (Starship).

\subsection{Beyond Semantic Relevance: Systematic failure of modern IR, NLI, and selective-prediction baselines}
\begin{figure}[t]
\centering
\includegraphics[width=\textwidth]{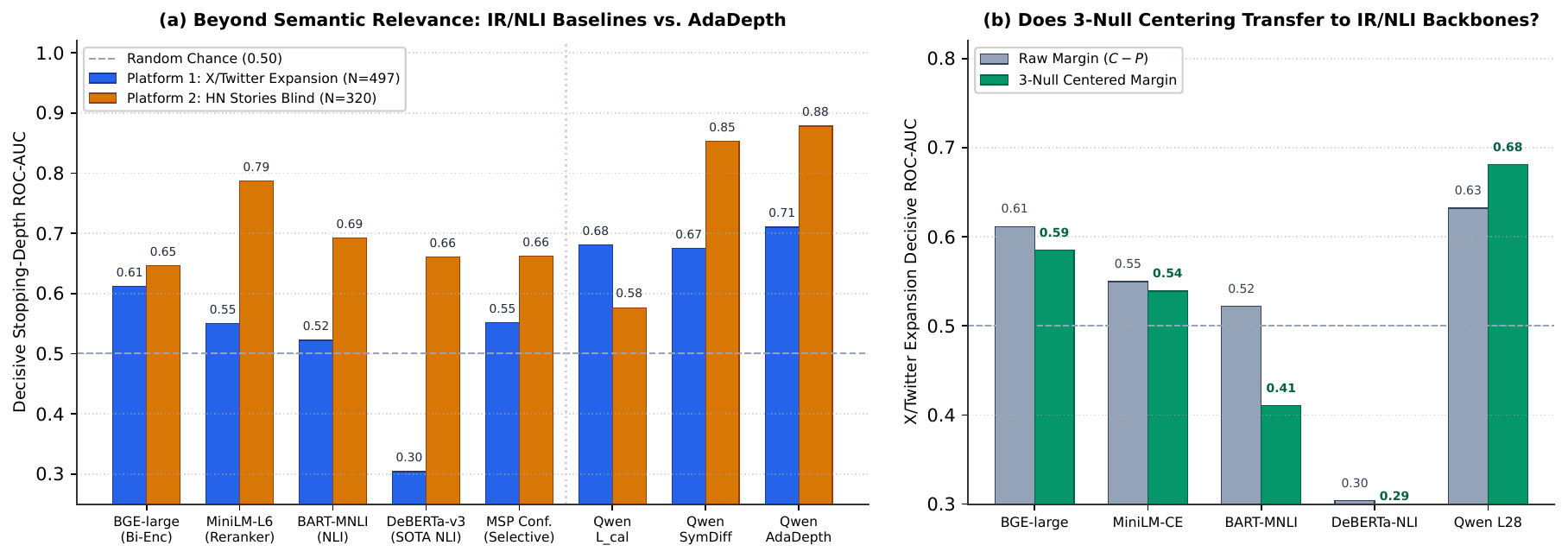}
\caption{Empirical proof of \emph{Beyond Semantic Relevance} and 3-null transferability. \textbf{(a)}~Decisive stopping-depth ROC-AUC across both Platform~1 (X/Twitter Expansion, $N=497$, blue) and Platform~2 (HN Stories Blind, $N=320$, orange): modern IR (\texttt{BGE-large}, \texttt{MiniLM-L6}), NLI (\texttt{BART-MNLI}; \texttt{DeBERTa-v3-large} inverting to 0.30~AUC on X/Twitter), and selective \texttt{MSP} all fall well behind \texttt{Qwen} \texttt{AdaDepth} (0.71~AUC on X/Twitter; 0.88~AUC on HN Stories). \textbf{(b)}~Effect of 3-null contextual centering on X/Twitter Expansion ROC-AUC: subtracting content-free null offsets degrades dedicated IR and NLI backbones (\texttt{BGE-large}: $0.61\to 0.59$; \texttt{MiniLM-CE}: $0.55\to 0.54$; \texttt{BART-MNLI}: $0.52\to 0.41$; \texttt{DeBERTa-NLI}: $0.30\to 0.29$), whereas final-layer LLM hidden-state margins (\texttt{Qwen L28}) gain $+5$~pp ($0.63\to 0.68$).}
\label{fig:ir_nli}
\Description{Two-panel bar chart. Panel (a) compares Decisive ROC-AUC across X/Twitter Expansion and HN Stories Blind for BGE-large, MiniLM-L6, BART-MNLI, DeBERTa-v3, MSP Confidence, Qwen L_cal, Qwen SymDiff, and Qwen AdaDepth. Panel (b) compares raw versus 3-null centered X/Twitter Expansion ROC-AUC for BGE-large, MiniLM-CE, BART-MNLI, DeBERTa-NLI, and Qwen L28, showing gains exclusively on Qwen L28.}
\end{figure}

\Cref{tab:tois_master} and \Cref{fig:ir_nli} test whether state-of-the-art semantic retrievers, cross-encoder rerankers, NLI classifiers, selective-prediction uncertainty scorers, or hierarchical hedging algorithms can distinguish evidence-supported Atomic links from over-descends.
Across all six baseline families, none achieves reliable gains over All-Backoff:

\begin{table}[t]
\centering
\caption{Master comparison of modern IR, NLI, Selective-Prediction, and Hierarchical-Hedging baselines against \texttt{AdaDepth} across held-out X/Twitter Expansion-497 ($|\mathcal{C}_{\mathrm{leaf}}|=299$, All-BO $=0.4789$) and blind HN Stories Full-320 ($|\mathcal{C}_{\mathrm{leaf}}|=150$, All-BO $=0.4969$). $\Delta$HEM is measured against unconditional All-Backoff; bold indicates best deployable result per column.}
\label{tab:tois_master}
\footnotesize
\setlength{\tabcolsep}{3.0pt}
\resizebox{\textwidth}{!}{%
\begin{tabular}{llcccccccccccc}
\toprule
& & \multicolumn{6}{c}{\textbf{X-Expansion-497 (All-BO $=0.4789$, Oracle $=0.6942$)}} & \multicolumn{6}{c}{\textbf{HN-Stories-320 (All-BO $=0.4969$, Oracle $=0.5469$)}} \\
\cmidrule(lr){3-8}\cmidrule(lr){9-14}
Family & Method / Signal & AllAUC & InPair & AURC$\downarrow$ & HEM$\uparrow$ & $\Delta$BO & Ret / OD & AllAUC & InPair & AURC$\downarrow$ & HEM$\uparrow$ & $\Delta$BO & Ret / OD \\
\midrule
Lexical IR & BM25 Margin ($C-P$) & 0.531 & 0.536 & -- & 0.461 & $-1.81$ & 79 / 101 & 0.674 & 0.860 & -- & 0.453 & $-4.38$ & 7 / 32 \\
Supervised & LOSO Logistic Reg. & 0.369 & -- & -- & 0.411 & $-6.84$ & -- & -- & -- & -- & -- & -- & -- \\
\midrule
\multirow{4}{*}{NLI}
 & DeBERTa-v3-large [raw] & 0.304 & 0.438 & 0.753 & 0.396 & $-8.25^{\dagger}$ & 16 / 63 & 0.661 & 0.485 & 0.842 & 0.453 & $-4.38^{*}$ & 7 / 43 \\
 & DeBERTa-v3-large [3-null] & 0.289 & 0.438 & 0.719 & 0.384 & $-9.46^{\dagger}$ & 38 / 93 & 0.651 & 0.485 & 0.864 & 0.472 & $-2.50$ & 5 / 28 \\
 & BART-large-MNLI [raw] & 0.522 & 0.476 & 0.584 & 0.487 & $+0.80$ & 40 / 41 & 0.692 & 0.544 & 0.813 & 0.475 & $-2.19$ & 4 / 13 \\
 & BART-large-MNLI [3-null] & 0.411 & 0.476 & 0.683 & 0.443 & $-3.62$ & 87 / 117 & 0.503 & 0.544 & 0.889 & 0.425 & $-7.19^{\dagger}$ & 10 / 47 \\
\midrule
\multirow{4}{*}{Dense / CE IR}
 & BGE-large Bi-Enc. [raw] & 0.612 & \textbf{0.685} & 0.529 & 0.435 & $-4.43$ & 101 / 136 & 0.647 & 0.779 & 0.837 & 0.406 & $-9.06^{\dagger}$ & 15 / 80 \\
 & BGE-large Bi-Enc. [3-null] & 0.585 & \textbf{0.685} & 0.554 & 0.455 & $-2.41$ & 85 / 107 & 0.709 & 0.779 & 0.807 & 0.416 & $-8.12^{\dagger}$ & 13 / 72 \\
 & MiniLM-L6 Cross-Enc. [raw] & 0.550 & 0.603 & 0.602 & 0.435 & $-4.43$ & 83 / 117 & 0.787 & 0.794 & 0.773 & 0.466 & $-3.12$ & 14 / 36 \\
 & MiniLM-L6 Cross-Enc. [3-null] & 0.539 & 0.603 & 0.604 & 0.455 & $-2.41$ & 79 / 99 & 0.764 & 0.794 & 0.781 & 0.478 & $-1.87$ & 14 / 31 \\
\midrule
\multirow{2}{*}{Selective}
 & MSP Confidence (Qwen) & 0.552 & 0.399 & 0.648 & 0.449 & $-3.02$ & 97 / 125 & 0.662 & 0.816 & 0.878 & 0.384 & $-11.25^{\dagger}$ & 15 / 85 \\
 & Neg. Entropy (Qwen) & 0.552 & 0.398 & 0.647 & 0.445 & $-3.42$ & 90 / 120 & 0.667 & 0.824 & 0.878 & 0.384 & $-11.25^{\dagger}$ & 15 / 85 \\
Hier. Hedge & DARTS Hedging (Qwen) & 0.568 & 0.527 & 0.549 & 0.479 & $+0.00$ & 45 / 49 & \textbf{0.891} & \textbf{0.956} & \textbf{0.719} & 0.484 & $-1.25$ & 14 / 21 \\
\midrule
\textbf{Ours (7B)} & \textbf{Qwen2.5-7B: \texttt{AdaDepth}} & \textbf{0.710} & 0.619 & \textbf{0.503} & \textbf{0.533} & $\mathbf{+5.43^{*}}$ & 70 / 47 & 0.878 & 0.897 & 0.734 & \textbf{0.503} & $\mathbf{+0.63}$ & 9 / 9 \\
\textbf{Ours (3.8B)} & \textbf{Phi-4-mini: \texttt{AdaDepth}} & 0.658 & 0.607 & 0.534 & 0.513 & $\mathbf{+3.42}$ & 53 / 41 & 0.787 & 0.912 & 0.773 & 0.500 & $\mathbf{+0.31}$ & 7 / 7 \\
\bottomrule
\end{tabular}%
}
\par\vspace{0.25em}\raggedright\scriptsize $\Delta$BO in percentage points (pp). Exact two-sided McNemar test vs.\ All-Backoff: $^{*}p<0.05$, $^{\dagger}p<0.001$. Max Ret / OD: 107 / 138 (X-Exp), 16 / 97 (HN-320).
\end{table}

\begin{enumerate}[leftmargin=*,nosep]
  \item \textbf{Severe NLI Inversion on Social Discourse:} \texttt{DeBERTa-v3-large} NLI achieves an inverted \textbf{0.304~AUC} on X-Expansion ($-8.25$~pp HEM vs.\ All-Backoff, $p=2\times 10^{-6}$). Because longer, detailed Parent Episode cards contain more multi-clause scope qualifiers that NLI models treat as harder-to-entail hypotheses than concise Atomic cards, they systematically assign higher entailment margins to unsupported leaves. Even on concise HN headlines, \texttt{DeBERTa-v3-large} (0.661~AUC, 0.453~HEM, $-4.38$~pp) and \texttt{BART-large-MNLI} (0.692~AUC, 0.475~HEM, $-2.19$~pp) both lose to All-Backoff, and 3-null centering degrades both NLI encoders further (replicating our OOD refusal finding on classification heads).
  \item \textbf{Semantic Similarity Confuses Entity Overlap with Evidential Sufficiency:} \texttt{BGE-large} achieves respectable within-pair ranking on X-Expansion (0.685~InPairAUC) and HN Stories (0.779), yet collapses at deployment: because Atomic cards carry salient named entities (e.g., ``SpaceX'', ``White Sands''), raw cosine similarity favors the Atomic card on nearly every post, leaving \textbf{136/138} over-descends on X-Expansion (0.435~HEM, $-4.43$~pp) and \textbf{80/97} on HN Stories (0.406~HEM, $-9.06$~pp, $p=0.0002$). Applying 3-null centering to \texttt{BGE-large} lifts HN AUC from 0.647 to 0.709 but still leaves 72/97 over-descends ($-8.12$~pp vs.\ All-Backoff). \texttt{MiniLM-L6} cross-encoder ranks strongly on HN headlines (0.787~AUC) yet still loses to All-Backoff on both X-Expansion ($-4.43$~pp) and HN Stories ($-3.12$~pp).
  \item \textbf{Selective Uncertainty and DARTS Fail Without Base-Rate Anchoring:} MSP confidence and negative entropy exhibit near-random within-pair discrimination on X-Expansion (0.399~InPairAUC) and keep 85/97 over-descends on HN Stories (0.384~HEM, $-11.25$~pp, $p=10^{-5}$). DARTS hierarchical hedging inherits \texttt{Direct\_AB}'s strong ranking on HN (0.891~AUC), yet without base-rate prevalence calibration it still over-keeps 21/97 over-descends on HN ($-1.25$~pp vs.\ All-Backoff) and ties All-Backoff on X-Expansion ($+0.00$~pp).
\end{enumerate}

\section{Results II: \texorpdfstring{\texttt{AdaDepth}}{AdaDepth}, Risk--Coverage, Accuracy--Specificity, and Efficiency}\label{sec:res_part2}

\subsection{\texorpdfstring{\texttt{AdaDepth}}{AdaDepth} wins 8/8 dataset--model cells over All-Backoff}
\Cref{tab:adadepth_ablation} compares \texttt{AdaDepth} against its individual components ($L_{\mathrm{cal}}$, \texttt{SymDiff}, \texttt{Direct\_AB}) and the fixed-weight ablation (\texttt{Fixed-0.5} + $\hat{\pi}_L$) across all four Gold evaluation sets and both LLM architectures (8 cells total).
\textbf{\texttt{AdaDepth} outperforms All-Backoff in all 8 cells:}
\begin{itemize}[leftmargin=*,nosep]
  \item \textbf{X-Expansion-497:} \texttt{Qwen2.5-7B} \texttt{AdaDepth} reaches \textbf{0.7101~AllAUC} (beating every single signal and every IR/NLI baseline), \textbf{58/107 Ret@25\%}, and \textbf{0.5332~HEM} ($+5.43$~pp over All-Backoff, McNemar $p=0.0141$; $+9.05$~pp over Orig \texttt{EviLink-G}, $p=4.5\times 10^{-5}$), cutting over-descending errors by 66\% ($138\to 47$) while retaining 70/107 true Atomic links. \texttt{Phi-4-mini} \texttt{AdaDepth} likewise reaches 0.6582~AllAUC and 0.5131~HEM ($+3.42$~pp vs.\ All-BO; $+7.04$~pp vs.\ Orig, $p=0.0043$).
  \item \textbf{X-Pilot-200 (High Atomic Density):} Whereas All-Backoff lowers Pilot HEM to 0.6750 ($-7.50$~pp), \texttt{AdaDepth} preserves high accuracy on both models (\textbf{0.7450~HEM}, $+7.00$~pp over All-Backoff; $p=0.0161$ on Qwen, $p=0.0043$ on Phi-4), reducing AURC from 0.286/0.464 ($L_{\mathrm{cal}}$) to \textbf{0.197/0.203} and boosting \texttt{Phi-4-mini} AUC by $+42.9$~pp over $L_{\mathrm{cal}}$ ($0.223\to 0.652$) and $+19.6$~pp over \texttt{Fixed-0.5} ($0.457\to 0.652$).
  \item \textbf{HN-Stories Blind CP1-320 and Audited Dual-160:} On blind HN-320, \texttt{AdaDepth} achieves \textbf{0.8782~AllAUC}, \textbf{15/16 Ret@25\%}, and \textbf{0.5031~HEM} on \texttt{Qwen2.5-7B} ($+0.63$~pp vs.\ All-BO; $+13.75$~pp vs.\ Orig, $p=2.4\times 10^{-9}$; reducing OD from $97\to 9$, $>90\%$ cut) and \textbf{0.5000~HEM} on \texttt{Phi-4-mini} ($+0.31$~pp vs.\ All-BO; $+13.44$~pp vs.\ Orig, $p=1.8\times 10^{-8}$; \texttt{Direct\_AB} scores 0.3688 in the same setting, whereas \texttt{AdaDepth} scores 0.5000). On fully dual-audited \texttt{HN-Dual-160}, both models reach \textbf{0.5000~HEM} ($+1.25$~pp over All-Backoff).
  \item \textbf{Ablation vs.\ \texttt{Fixed-0.5}:} Replacing ontology-intrinsic $w_{\mathrm{gap}}(e,a)$ with a constant $0.5$ weight lowers HEM by $0.9$~pp on \texttt{Qwen} HN-320 ($0.503\to 0.494$, AUC $0.878\to 0.807$) and $2.5$~pp on \texttt{Phi-4} Pilot ($0.745\to 0.720$, AUC $0.652\to 0.457$), while raising \texttt{Qwen} X-Exp by $0.6$~pp ($0.533\to 0.539$).
\end{itemize}

\begin{table}[t]
\centering
\caption{Zero-parameter \texttt{AdaDepth} ablation across all four Gold evaluation sets and two open-weight LLM families (8/8 cells beating All-Backoff). Bold marks best deployable HEM per block; $\Delta$BO in percentage points.}
\label{tab:adadepth_ablation}
\footnotesize
\setlength{\tabcolsep}{2.8pt}
\resizebox{\textwidth}{!}{%
\begin{tabular}{llcccccccccccc}
\toprule
& & \multicolumn{3}{c}{\textbf{X-Exp-497} (BO 0.479)} & \multicolumn{3}{c}{\textbf{X-Pilot-200} (BO 0.675)} & \multicolumn{3}{c}{\textbf{HN-CP1-320} (BO 0.497)} & \multicolumn{3}{c}{\textbf{HN-Dual-160} (BO 0.488)} \\
\cmidrule(lr){3-5}\cmidrule(lr){6-8}\cmidrule(lr){9-11}\cmidrule(lr){12-14}
Model & Method & AUC & AURC & HEM ($\Delta$BO) & AUC & AURC & HEM ($\Delta$BO) & AUC & AURC & HEM ($\Delta$BO) & AUC & AURC & HEM ($\Delta$BO) \\
\midrule
\multirow{5}{*}{Qwen2.5-7B}
 & $L_{\mathrm{cal}}$ (Path B) & 0.681 & 0.529 & 0.519 ($+4.0$) & 0.484 & 0.286 & 0.740 ($+6.5$) & 0.577 & 0.872 & 0.478 ($-1.9$) & 0.731 & 0.867 & 0.481 ($-0.6$) \\
 & SymDiff (Path C) & 0.675 & 0.503 & 0.507 ($+2.8$) & 0.660 & 0.206 & 0.715 ($+4.0$) & 0.853 & 0.772 & 0.500 ($+0.3$) & 0.894 & 0.771 & 0.494 ($+0.6$) \\
 & Direct\_AB (Path C) & 0.568 & 0.549 & 0.477 ($-0.2$) & 0.644 & 0.215 & 0.735 ($+6.0$) & 0.891 & 0.719 & 0.478 ($-1.9$) & 0.872 & 0.803 & 0.469 ($-1.9$) \\
 & Fixed-0.5 + $\hat{\pi}_L$ & 0.707 & 0.485 & \textbf{0.539} ($+6.0$) & 0.598 & 0.227 & 0.740 ($+6.5$) & 0.807 & 0.766 & 0.494 ($-0.3$) & 0.923 & 0.785 & \textbf{0.500} ($+1.3$) \\
 & \textbf{\texttt{AdaDepth} (Ours)} & \textbf{0.710} & 0.503 & 0.533 ($\mathbf{+5.4}$) & \textbf{0.652} & \textbf{0.197} & \textbf{0.745} ($\mathbf{+7.0}$) & \textbf{0.878} & \textbf{0.734} & \textbf{0.503} ($\mathbf{+0.6}$) & 0.891 & 0.784 & \textbf{0.500} ($\mathbf{+1.3}$) \\
\midrule
\multirow{5}{*}{Phi-4-mini}
 & $L_{\mathrm{cal}}$ (Path B) & 0.628 & 0.574 & 0.517 ($+3.8$) & 0.223 & 0.464 & 0.710 ($+3.5$) & 0.486 & 0.904 & 0.478 ($-1.9$) & 0.635 & 0.905 & 0.469 ($-1.9$) \\
 & SymDiff (Path C) & 0.589 & 0.577 & 0.477 ($-0.2$) & 0.728 & 0.195 & 0.690 ($+1.5$) & 0.793 & 0.807 & 0.497 ($+0.0$) & 0.750 & 0.890 & 0.488 ($+0.0$) \\
 & Direct\_AB (Path C) & 0.606 & 0.529 & 0.445 ($-3.4$) & 0.582 & 0.254 & \textbf{0.750} ($+7.5$) & 0.877 & 0.738 & 0.369 ($-12.8$) & 0.926 & 0.762 & 0.363 ($-12.5$) \\
 & Fixed-0.5 + $\hat{\pi}_L$ & 0.657 & 0.529 & \textbf{0.523} ($+4.4$) & 0.457 & 0.304 & 0.720 ($+4.5$) & 0.688 & 0.822 & 0.494 ($-0.3$) & 0.846 & 0.820 & 0.494 ($+0.6$) \\
 & \textbf{\texttt{AdaDepth} (Ours)} & \textbf{0.658} & 0.534 & 0.513 ($\mathbf{+3.4}$) & \textbf{0.652} & \textbf{0.203} & 0.745 ($\mathbf{+7.0}$) & \textbf{0.787} & \textbf{0.773} & \textbf{0.500} ($\mathbf{+0.3}$) & 0.872 & 0.799 & \textbf{0.500} ($\mathbf{+1.3}$) \\
\bottomrule
\end{tabular}%
}
\end{table}

\subsection{Prospective Stage-2 Blind Validation (\texorpdfstring{\texttt{Reserve-160}}{Reserve-160}) and Full \texorpdfstring{\texttt{HN-All-480}}{HN-All-480} Results}\label{sec:stage2_res}
After confirming exact file consistency across all 12 pre-annotation prediction files, evaluating the dual-adjudicated Gold labels of \texttt{Reserve-160} ($|\mathcal{C}_{\mathrm{leaf}}|=72$, $\mathrm{KeepOK}=8$, $\mathrm{OD}_{\mathrm{raw}}=54$, $\mathrm{BackoffOK}=34$) and its fully dual-audited subset \texttt{Reserve-Dual-80} ($N=80$, $|\mathcal{C}_{\mathrm{leaf}}|=34$, $\mathrm{KeepOK}=5$, $\mathrm{OD}_{\mathrm{raw}}=24$, $\mathrm{BackoffOK}=11$) provides a prospective test of frozen \texttt{AdaDepth v1.0} (\Cref{tab:stage2_blind}).
Three findings emerge:
\begin{enumerate}[leftmargin=*,nosep]
  \item \textbf{Replication of the All-Backoff Paradox and Baseline Failure:} On prospective \texttt{Reserve-160}, the All-Backoff Paradox reproduces cleanly (\texttt{Orig} $0.4125 \to$ \texttt{All-Backoff} $0.5750$, $+16.25$~pp; Oracle $0.6250$), and \emph{every modern IR, NLI, Selective, and DARTS baseline underperforms All-Backoff}, including \texttt{BGE-large} ($-9.37$~pp raw, $p=0.011$; $-6.25$~pp cal), \texttt{MiniLM-L6} ($-5.00$~pp), \texttt{DeBERTa-v3-large} ($-3.75$~pp), \texttt{BART-MNLI} ($-3.75$~pp raw / $-6.87$~pp cal, $p=0.019$), \texttt{MSP} ($-13.75$~pp, $p=0.0005$), and \texttt{DARTS} ($-0.62$~pp). Aggregated across the complete 480-item Hacker News corpus (\texttt{HN-All-480}, $|\mathcal{C}_{\mathrm{leaf}}|=222$, $\mathrm{All\text{-}BO}=0.5229$), every relevance, NLI, selective, DARTS, and \texttt{Fixed-0.5} comparator falls below All-Backoff (with statistically significant losses for \texttt{BGE-large} $p<10^{-5}$, \texttt{MSP} $p<10^{-7}$, \texttt{DeBERTa} $p=0.0015$, and \texttt{MiniLM-L6} $p=0.025$).
  \item \textbf{Both Models Beat All-Backoff on Dual-Audited \texttt{Reserve-Dual-80}, and \texttt{Phi-4-mini} Leads Raw-Threshold Baselines:} On the fully dual-reviewed \texttt{Reserve-Dual-80} subset (Orig $0.4250$, All-Backoff $0.5000$, Oracle $0.5625$), \emph{both} frozen \texttt{AdaDepth v1.0} models beat All-Backoff by \textbf{$+3.75$~pp} (\textbf{0.5375~HEM}, $+11.25$~pp vs.\ Orig, $p=0.0225$) with \textbf{1.0000~InPairAUC} while reducing OD by 91.7--95.8\% (\texttt{Qwen}: $24\to 2$; \texttt{Phi-4}: $24\to 1$), whereas \texttt{BGE-large} ($-2.50$~pp), \texttt{MiniLM-L6} ($-1.25$~pp), \texttt{MSP} ($-7.50$~pp), and \texttt{Phi-4 Direct\_AB} ($-6.25$~pp) all lose to All-Backoff. Across full \texttt{Reserve-160} and full \texttt{HN-All-480}, frozen \texttt{Phi-4-mini AdaDepth} achieves the \textbf{highest HEM among all 21 raw-threshold methods}: \textbf{0.5875~HEM} on \texttt{Reserve-160} ($+1.25$~pp over All-BO, $+17.50$~pp over Orig, $p=1.9\times 10^{-6}$; highest \texttt{InPairAUC} at \textbf{0.8571}; 94.4\% OD cut $54\to 3$) and \textbf{0.5292~HEM} ($254/480$) on \texttt{HN-All-480} ($+0.62$~pp over All-BO, $+14.79$~pp over Orig, $p=6.9\times 10^{-14}$; 93.4\% OD cut $151\to 10$), eliminating \texttt{Direct\_AB}'s raw logit collapse ($0.4188\to 0.5875$ on Reserve; $0.3854\to 0.5292$ on HN-480).
  \item \textbf{Conservative Interpretation Under Sparse HN Leaf Prevalence:} Across Hacker News, true \texttt{KeepOK} leaves are sparse ($n=16$ in CP1, $n=8$ in Reserve, $n=24$ combined), capping the entire achievable headroom from All-Backoff to Oracle at only 5.0~pp. Consequently, absolute HEM gains over All-Backoff on HN splits are small ($+0.31$ to $+1.25$~pp, corresponding to $+1$ to $+3$ net items, individually non-significant), so HN evidence rests primarily on strong ROC-AUC, low Risk--Coverage AURC, $>90\%$ over-descending reduction, and eliminating cross-model logit drift while preserving 46--50\% of true Atomic links rather than collapsing to 0\% leaf specificity. On \texttt{Reserve-160} alone, frozen \texttt{Qwen2.5-7B} \texttt{AdaDepth} lifts HEM by $+15.00$~pp over the base generator ($0.5625$, $p=3.9\times 10^{-5}$, 87.0\% OD cut $54\to 7$) and matches All-Backoff across all 480 HN items ($251/480 = 0.5229$~HEM, $+14.17$~pp vs.\ Orig, $p=1.7\times 10^{-13}$, 0.805~AllAUC, 0.746~AURC), while trailing All-Backoff on Reserve by 2 net items ($-1.25$~pp, $p=0.727$) because \texttt{SymDiff} weakened on the 8 positive items (0.682~AUC vs.\ 0.794 for \texttt{Direct\_AB}). By contrast, on higher-density X/Twitter splits, \texttt{AdaDepth} achieves statistically significant HEM gains over All-Backoff ($p=0.014$ on \texttt{X-Exp Qwen}; $p=0.016/0.0043$ on \texttt{X-Pilot}).
\end{enumerate}

\begin{table}[t]
\centering
\caption{Prospective Stage-2 Blind validation on frozen, pre-annotation consistency-checked \texttt{HN-Reserve-160} ($N=160$, $|\mathcal{C}_{\mathrm{leaf}}|=72$, $\mathrm{KeepOK}=8$, $\mathrm{OD}_{\mathrm{raw}}=54$; Orig $=0.4125$, All-BO $=0.5750$, Oracle $=0.6250$), its dual-audited subset \texttt{Reserve-Dual-80} ($N=80$, All-BO $=0.5000$), and combined \texttt{HN-All-480} ($N=480$, $|\mathcal{C}_{\mathrm{leaf}}|=222$, $\mathrm{KeepOK}=24$, $\mathrm{OD}_{\mathrm{raw}}=151$; Orig $=0.3812$, All-BO $=0.5229$, Oracle $=0.5729$). Every \texttt{HN-All-480} \texttt{AdaDepth} decision strictly concatenates separate frozen CP1 and Reserve decisions with zero cross-split pooling.}
\label{tab:stage2_blind}
\footnotesize
\setlength{\tabcolsep}{2.6pt}
\resizebox{\textwidth}{!}{%
\begin{tabular}{llccccccccccccccc}
\toprule
& & \multicolumn{6}{c}{\textbf{Prospective Reserve-160 (BO 0.575)}} & \multicolumn{3}{c}{\textbf{Dual-80 (0.500)}} & \multicolumn{6}{c}{\textbf{Combined HN-All-480 (BO 0.523)}} \\
\cmidrule(lr){3-8}\cmidrule(lr){9-11}\cmidrule(lr){12-17}
Family & Method / Signal & AUC & InPair & AURC$\downarrow$ & HEM$\uparrow$ & $\Delta$BO & Ret / OD & AUC & InPair & HEM & AUC & InPair & AURC$\downarrow$ & HEM$\uparrow$ & $\Delta$BO & Ret / OD \\
\midrule
\multirow{2}{*}{NLI}
 & DeBERTa-v3-large [raw] & 0.603 & 0.619 & 0.902 & 0.538 & $-3.75$ & 3 / 24 & 0.673 & 0.600 & 0.513 & 0.649 & 0.402 & 0.868 & 0.479 & $-4.38^{*}$ & 10 / 73 \\
 & BART-large-MNLI [raw] & 0.658 & 0.619 & 0.839 & 0.538 & $-3.75$ & 3 / 12 & 0.855 & \textbf{1.000} & 0.525 & 0.719 & 0.669 & 0.798 & 0.508 & $-1.46$ & 13 / 28 \\
\midrule
\multirow{2}{*}{IR}
 & BGE-large Bi-Enc. [raw] & 0.732 & 0.714 & 0.764 & 0.481 & $-9.37^{*}$ & 8 / 34 & 0.746 & 0.800 & 0.475 & 0.674 & 0.746 & 0.813 & 0.431 & $-9.17^{\dagger}$ & 23 / 114 \\
 & MiniLM-L6 Cross-Enc. [raw] & 0.814 & 0.762 & 0.750 & 0.525 & $-5.00$ & 6 / 20 & 0.818 & 0.800 & 0.488 & 0.796 & 0.787 & 0.760 & 0.485 & $-3.75^{*}$ & 20 / 56 \\
\midrule
Selective & MSP Confidence (Qwen) & 0.574 & 0.476 & 0.878 & 0.438 & $-13.75^{\dagger}$ & 8 / 47 & 0.709 & 0.800 & 0.425 & 0.629 & 0.681 & 0.879 & 0.402 & $-12.08^{\dagger}$ & 23 / 132 \\
Hier. Hedge & DARTS Hedging (Qwen) & 0.794 & 0.714 & 0.749 & 0.569 & $-0.62$ & 6 / 9 & \textbf{1.000} & \textbf{1.000} & \textbf{0.550} & \textbf{0.861} & \textbf{0.873} & \textbf{0.724} & 0.513 & $-1.04$ & 20 / 30 \\
\midrule
\multirow{4}{*}{Qwen2.5-7B}
 & Direct\_AB (Path C) & 0.794 & 0.714 & 0.749 & 0.550 & $-2.50$ & 6 / 13 & \textbf{1.000} & \textbf{1.000} & \textbf{0.550} & \textbf{0.861} & \textbf{0.873} & \textbf{0.724} & 0.502 & $-2.08$ & 21 / 41 \\
 & SymDiff (Path C) & 0.682 & 0.714 & 0.795 & 0.575 & $+0.00$ & 0 / 0 & 0.836 & \textbf{1.000} & 0.500 & 0.791 & 0.769 & 0.777 & 0.525 & $+0.21$ & 1 / 0 \\
 & Fixed-0.5 + $\hat{\pi}_L$ & 0.757 & 0.762 & 0.751 & 0.569 & $-0.62$ & 4 / 6 & 0.800 & \textbf{1.000} & 0.538 & 0.789 & 0.805 & 0.758 & 0.519 & $-0.42$ & 11 / 17 \\
 & \textbf{\texttt{AdaDepth v1.0} (Frozen)} & 0.658 & 0.714 & 0.775 & 0.563 & $-1.25$ & 3 / 7 & 0.800 & \textbf{1.000} & 0.538 & 0.805 & 0.834 & 0.746 & 0.523 & $+0.00$ & 12 / 16 \\
\midrule
\multirow{4}{*}{Phi-4-mini}
 & Direct\_AB (Path C) & \textbf{0.820} & 0.810 & \textbf{0.712} & 0.419 & $-15.62^{\dagger}$ & 8 / 53 & 0.982 & 0.900 & 0.438 & 0.856 & 0.828 & 0.729 & 0.385 & $-13.75^{\dagger}$ & 24 / 149 \\
 & SymDiff (Path C) & 0.691 & 0.714 & 0.827 & 0.575 & $+0.00$ & 0 / 0 & 0.873 & \textbf{1.000} & 0.500 & 0.757 & 0.808 & 0.812 & 0.523 & $+0.00$ & 0 / 0 \\
 & Fixed-0.5 + $\hat{\pi}_L$ & 0.699 & 0.810 & 0.772 & 0.569 & $-0.62$ & 3 / 4 & 0.709 & \textbf{1.000} & 0.525 & 0.692 & 0.781 & 0.810 & 0.519 & $-0.42$ & 9 / 12 \\
 & \textbf{\texttt{AdaDepth v1.0} (Frozen)} & 0.607 & \textbf{0.857} & 0.803 & \textbf{0.588} & $\mathbf{+1.25}$ & 4 / 3 & 0.746 & \textbf{1.000} & 0.538 & 0.727 & 0.846 & 0.782 & \textbf{0.529} & $\mathbf{+0.62}$ & 11 / 10 \\
\bottomrule
\end{tabular}%
}
\par\vspace{0.25em}\raggedright\scriptsize $\Delta$BO in percentage points. Exact two-sided McNemar test vs.\ All-Backoff: $^{*}p<0.05$, $^{\dagger}p<0.001$. Both frozen \texttt{AdaDepth v1.0} models improve over Base Generator at $p<4\times 10^{-5}$ on \texttt{Reserve-160}, $p=0.0225$ on \texttt{Reserve-Dual-80}, and $p<2\times 10^{-13}$ on \texttt{HN-All-480}. Note that \texttt{SymDiff} matches All-BO only by keeping 0/8 on Reserve and 0--1/24 on HN-480, whereas \texttt{AdaDepth} preserves 46--50\% of true Atomic links.
\end{table}

\subsection{Two-Factor Pivot Ablation: Prevalence Anchoring (\texorpdfstring{$\hat{\pi}_L$}{pi\_L}) vs.\ Gap-Aware Fusion}\label{sec:c2_pivot}
A critical question for any composite calibrator is credit assignment: \emph{do \texttt{AdaDepth}'s gains stem primarily from its unlabeled prevalence threshold $\hat{\pi}_L$, or does Granularity-Gap-aware fusion $S_{\mathrm{AdaDepth}}$ add cross-regime value beyond thresholding?}
To isolate both factors cleanly, \Cref{tab:c2_pivot} equips \emph{every} external IR, NLI, selective, and prompt baseline with \texttt{AdaDepth}'s exact same unlabeled quantile gate $\hat{\tau} = Q_{1-\hat{\pi}_L}(\cdot)$ across all 8 benchmark cells (4 datasets $\times$ 2 backbone $\hat{\pi}_L$ anchors).

Two conclusions emerge from this decomposition:
\begin{enumerate}[leftmargin=*,nosep]
  \item \textbf{Factor~1 (Unlabeled $\hat{\pi}_L$ Anchoring Improves IR and Prompt Scorers):} Transplanting our unlabeled 3-null prevalence threshold improves the IR and prompt-log-odds baselines (\texttt{BGE-large} moves from $-4.43/-9.06$~pp raw to $+1.81/-1.56$~pp, \texttt{MiniLM-L6} reaches $+1.97$~pp mean $\Delta$BO with 6/8 wins, and \texttt{Direct\_AB/DARTS} on \texttt{Phi-4-mini} moves from $-12.8/-15.6$~pp logit drift to $+1.56/+1.25$~pp on HN Stories), while the NLI encoders stay below All-Backoff on \texttt{X-Exp} ($-4.83$~pp for \texttt{DeBERTa-v3-large} on Qwen). As detailed in \Cref{app:robustness}, \texttt{Phi-4-mini}'s unlabeled $\hat{\pi}_L$ tracks the $6\times$ base-rate shift across all four datasets with \textbf{2.8~pp MAE} (0.67~pp error on both \texttt{X-Exp} and \texttt{HN-CP1}).
  \item \textbf{Factor~2 (Gap-Aware Fusion Adds $+0.45$~pp Mean Gain Beyond Prevalence Anchoring):} Under the same $\hat{\pi}_L$ threshold, \textbf{\texttt{AdaDepth} achieves the highest mean gain over All-Backoff across all 8 cells ($+2.97$~pp, 7/8 wins)}, adding $+0.45$~pp over the strongest single-signal comparator (\texttt{SymDiff + $\hat{\pi}_L$}, $+2.52$~pp, 5/8) and outperforming \texttt{Fixed-0.5 + $\hat{\pi}_L$} ($+2.45$~pp, 4/8), \texttt{MSP + $\hat{\pi}_L$} ($+2.31$~pp), \texttt{Direct\_AB/DARTS + $\hat{\pi}_L$} ($+2.18$~pp, 6/8), \texttt{MiniLM-L6 + $\hat{\pi}_L$} ($+1.97$~pp), \texttt{BGE-large + $\hat{\pi}_L$} ($+1.84$~pp), $L_{\mathrm{cal}}$ ($+1.61$~pp), \texttt{BART + $\hat{\pi}_L$} ($+1.24$~pp), and \texttt{DeBERTa-v3 + $\hat{\pi}_L$} ($+0.54$~pp). Auditing strict exact-$K$ uniform tie-breaking (to account for discrete prompt-logit ties at the quantile boundary) leaves the ranking unchanged: \texttt{AdaDepth} remains first at $+2.97$~pp vs.\ $+2.52$~pp (\texttt{SymDiff + $\hat{\pi}_L$}), $+2.45$~pp (\texttt{Fixed-0.5}), $+2.33$~pp (\texttt{MSP}), and $+2.13$~pp (\texttt{Direct\_AB/DARTS}).
  \item \textbf{Regime Trade-Off on High-Gap HN vs.\ Low-Gap X/Twitter and Empirical Check of Proposition~\ref{prop:quantile}:} Once equipped with our $\hat{\pi}_L$ anchor, pure prompt-side \texttt{Phi-4 Direct\_AB + $\hat{\pi}_L$} reaches $258/480 = 0.5375$~HEM on high-gap \texttt{HN-All-480} (inclusive $K=24$, $\mathrm{Net}=+7$; or $0.5361$ at strict $K=22$, $\mathrm{Net}=+6.34$), edging out \texttt{Phi-4 AdaDepth} ($254/480 = 0.5292$, $\mathrm{Net}=+3$) on HN alone, yet losing sharply on noisy low-gap \texttt{X-Exp} ($-0.20$~pp Qwen / $+1.81$~pp Phi-4 vs.\ \texttt{AdaDepth} $+5.43 / +3.42$~pp). Finally, \texttt{X-Exp Phi-4-mini} at fixed $K=109$ directly illustrates Proposition~\ref{prop:quantile}(ii)'s open-set substitution effect: \texttt{AdaDepth} retains more true leaves than $L_{\mathrm{cal}}$ ($k_1=53$ vs.\ $49$) while reducing kept upstream errors ($k_3=20$ vs.\ $30$), yet increases kept \texttt{BackoffOK} items ($k_2=36$ vs.\ $30$), yielding $\mathrm{Net}=k_1-k_2=+17$ vs.\ $+19$.
\end{enumerate}

\begin{table}[t]
\centering
\caption{Two-factor pivot ablation: $\Delta$HEM vs.\ unconditional All-Backoff (in pp) when \emph{every} baseline receives \texttt{AdaDepth}'s unlabeled prevalence threshold $\hat{\pi}_L$ across all 8 benchmark cells (4 splits $\times$ 2 backbones Q/P). Bold marks the best value in each column.}
\label{tab:c2_pivot}
\footnotesize
\setlength{\tabcolsep}{3.1pt}
\resizebox{\textwidth}{!}{%
\begin{tabular}{lcccccccccc}
\toprule
& \multicolumn{2}{c}{\textbf{X-Exp-497}} & \multicolumn{2}{c}{\textbf{X-Pilot-200}} & \multicolumn{2}{c}{\textbf{HN-CP1-320}} & \multicolumn{2}{c}{\textbf{HN-Reserve-160}} & \multicolumn{2}{c}{\textbf{8-Cell Summary}} \\
\cmidrule(lr){2-3}\cmidrule(lr){4-5}\cmidrule(lr){6-7}\cmidrule(lr){8-9}\cmidrule(lr){10-11}
Method (all using identical $\hat{\pi}_L$) & Qwen & Phi-4 & Qwen & Phi-4 & Qwen & Phi-4 & Qwen & Phi-4 & Mean $\Delta$BO & Wins $>0$ \\
\midrule
DeBERTa-v3-large + $\hat{\pi}_L$ & $-4.83$ & $-3.82$ & $+7.50$ & $\mathbf{+8.00}$ & $-0.31$ & $+0.31$ & $-1.25$ & $-1.25$ & $+0.54$ & 3 / 8 \\
BART-large-MNLI + $\hat{\pi}_L$ & $+0.40$ & $+1.61$ & $+7.50$ & $+6.00$ & $-0.63$ & $+0.00$ & $-3.12$ & $-1.87$ & $+1.24$ & 4 / 8 \\
$L_{\mathrm{cal}}$ ($\tau=0$, native $\hat{\pi}_L$) & $+4.02$ & $+3.82$ & $+6.50$ & $+3.50$ & $-1.87$ & $-1.87$ & $-1.25$ & $+0.00$ & $+1.61$ & 4 / 8 \\
BGE-large + $\hat{\pi}_L$ & $+1.81$ & $\mathbf{+4.63}$ & $+6.50$ & $+5.50$ & $-1.56$ & $-0.94$ & $-1.25$ & $+0.00$ & $+1.84$ & 4 / 8 \\
MiniLM-L6 CE + $\hat{\pi}_L$ & $+0.40$ & $+1.41$ & $+7.50$ & $+5.50$ & $+0.63$ & $+0.31$ & $\mathbf{+0.00}$ & $+0.00$ & $+1.97$ & 6 / 8 \\
Direct\_AB / DARTS + $\hat{\pi}_L$ & $-0.20$ & $+1.81$ & $+7.00$ & $+6.00$ & $+0.63$ & $\mathbf{+1.56}$ & $-0.62$ & $\mathbf{+1.25}$ & $+2.18$ & 6 / 8 \\
MSP Confidence + $\hat{\pi}_L$$^{a}$ & $+1.01$ & $+1.61$ & $\mathbf{+8.00}$ & $+6.00$ & $+0.31$ & $\mathbf{+1.56}$ & $-1.25$ & $\mathbf{+1.25}$ & $+2.31$ & \textbf{7 / 8} \\
Fixed-0.5 + $\hat{\pi}_L$ & $\mathbf{+6.04}$ & $+4.43$ & $+6.50$ & $+4.50$ & $-0.31$ & $-0.31$ & $-0.62$ & $-0.62$ & $+2.45$ & 4 / 8 \\
SymDiff + $\hat{\pi}_L$ & $+4.43$ & $+2.01$ & $+7.50$ & $+6.50$ & $\mathbf{+0.94}$ & $+0.00$ & $-0.62$ & $-0.62$ & $+2.52$ & 5 / 8 \\
\midrule
\textbf{\texttt{AdaDepth} ($w_{\mathrm{gap}}$ fusion + $\hat{\pi}_L$)} & $+5.43$ & $+3.42$ & $+7.00$ & $+7.00$ & $+0.63$ & $+0.31$ & $-1.25$ & $\mathbf{+1.25}$ & $\mathbf{+2.97}$ & \textbf{7 / 8} \\
\bottomrule
\end{tabular}%
}
\par\vspace{0.25em}\raggedright\scriptsize $^{a}$On \texttt{Phi-4-mini} HN splits, raw \texttt{Direct\_AB} logits are strictly positive ($P_B>0.5$ on $>98\%$ of items), so undirected MSP collapses to \texttt{Direct\_AB} and reproduces its exact HN score. Under strict exact-$K$ uniform tie-breaking, 8-cell mean $\Delta$BO remains identical for \texttt{AdaDepth} ($+2.97$~pp), \texttt{SymDiff} ($+2.52$), and \texttt{Fixed-0.5} ($+2.45$), and shifts slightly for \texttt{Direct\_AB/DARTS} ($+2.13$) and \texttt{MSP} ($+2.33$).
\end{table}

\subsection{Direct \texorpdfstring{$\Delta(e,a)$}{Delta(e,a)} Verification and Automatic Ontology Transfer Probe}\label{sec:direct_delta}
Two foundational questions remain: (1)~if we prompt the backbone \emph{directly} to verify whether post $p$ asserts the incremental claim $\Delta(e,a)$ as its main proposition under the exact same upstream proposals and budget $\hat{\pi}_L$, how does it compare with \texttt{AdaDepth}? And (2)~can $\Delta(e,a)$ be extracted \emph{automatically} from standard parent--child event descriptions $(c_e, c_a)$ without manual $\Delta(e,a)$ authoring?
Recall that \texttt{AdaDepth} itself never consumes manually authored $\Delta(e,a)$ or disqualifying-framing text on any dataset, using only standard \texttt{(parent\_desc, child\_desc)} summaries.
To test both questions directly on \texttt{Qwen2.5-7B-Instruct} across all 556 leaf candidates (46 unique card pairs), we run two zero-parameter \texttt{Direct\_Delta} first-token Yes/No log-odds baselines at matched budget $\hat{\pi}_L$:
(i)~\textbf{\texttt{Direct\_Delta (Auto-$\Delta$)}} (all 4 splits, zero manual $\Delta$): offline, \texttt{Qwen2.5-7B} reads only raw \texttt{(parent\_desc, child\_desc)} and generates a single concise incremental assertion $\widehat{\Delta}_{\mathrm{auto}}(e,a)$, then scores each post directly against $\widehat{\Delta}_{\mathrm{auto}}(e,a)$; and
(ii)~\textbf{\texttt{Direct\_Delta (Manual-$\Delta$)}} (\texttt{HN-CP1} and \texttt{HN-Reserve}): scores each headline directly against the human-authored $\Delta(e,a)$ and disqualifying parent framing.

\Cref{tab:direct_delta} reports exact per-split ranking, HEM, and $(k_1\text{ KeepOK}, k_2\text{ BackoffOK}, k_3\text{ UpstreamErr}, \mathrm{Net}=k_1-k_2)$ decompositions. Two sharp evaluation findings emerge:
\begin{enumerate}[leftmargin=*,nosep]
  \item \textbf{Direct $\Delta(e,a)$ Verification Excels on High-Gap HN and Transfers Without Manual $\Delta(e,a)$ Authoring:} On crisp high-gap Hacker News headlines, directly verifying manual $\Delta(e,a)$ at budget $\hat{\pi}_L$ reaches \textbf{0.950~AUC / 0.525~HEM} ($\mathrm{Net}=+9$) on \texttt{HN-CP1} and \textbf{0.879~AUC / 0.588~HEM} ($\mathrm{Net}=+2$) on prospective \texttt{HN-Reserve} ($262/480 = 0.5458$~HEM across \texttt{HN-All-480}, retaining $21/24$ true Atomic links). Notably, \textbf{fully automatic $\widehat{\Delta}_{\mathrm{auto}}(e,a)$} achieves \textbf{0.901~AUC} on \texttt{HN-CP1} ($\mathrm{Net}=+4$) and \textbf{0.873~AUC / 0.588~HEM} ($\mathrm{Net}=+2$, matching manual $\Delta$ identically) on prospective \texttt{HN-Reserve} ($257/480 = 0.5354$~HEM across \texttt{HN-All-480}), showing that operational $\Delta(e,a)$ boundaries can be extracted automatically from standard parent--child event cards without manual rubric authoring.
  \item \textbf{Prompt-Only $\Delta$-Verification Collapses on Noisy Low-Gap Social Discourse:} On held-out \texttt{X-Exp-497}, \texttt{Direct\_Delta (Auto-$\Delta$)} drops to \textbf{0.527~AUC} and \textbf{0.463~HEM} ($-1.61$~pp below All-Backoff; $k_1=55, k_2=63, \mathrm{Net}=-8$), trailing \texttt{AdaDepth} (\textbf{0.710~AUC}, \textbf{0.533~HEM}, $+5.43$~pp, $k_1=70, k_2=43, \mathrm{Net}=+27$) by \textbf{$+7.04$~pp ($+35$ net items)}, and also trails \texttt{AdaDepth} on \texttt{X-Pilot} AURC ($0.303$ vs.\ $0.197$). Thus, direct $\Delta$-verification is the strongest single specialist when card granularity gaps are sharp and text is headline-clean, whereas \texttt{AdaDepth} achieves higher cross-regime accuracy and ranking across noisy social media and headlines combined (4-split \texttt{Qwen} mean: \textbf{0.725~AUC}, \textbf{$+2.95$~pp} strict-$K$ $\Delta$BO vs.\ $0.704$~AUC, $+1.90$~pp for \texttt{Direct\_Delta (Auto-$\Delta$)}).
\end{enumerate}

\begin{table}[t]
\centering
\caption{Direct $\Delta(e,a)$ main-assertion verification and automatic $\widehat{\Delta}_{\mathrm{auto}}(e,a)$ transfer probe on \texttt{Qwen2.5-7B} at matched $\hat{\pi}_L$ budget across all four splits (plus summed \texttt{HN-All-480}), reporting exact kept counts $(k_1\text{ KeepOK}, k_2\text{ BackoffOK}, k_3\text{ UpstreamErr})$ and $\mathrm{Net} = k_1 - k_2$ (inclusive / strict exact-$K$ when ties occur).}
\label{tab:direct_delta}
\footnotesize
\setlength{\tabcolsep}{3.4pt}
\resizebox{\textwidth}{!}{%
\begin{tabular}{llccccccccc}
\toprule
Dataset (All-BO, $K$) & Method (all at Qwen $\hat{\pi}_L$) & AllAUC & AURC$\downarrow$ & Raw $\Delta$BO & HEM & $\Delta$BO & $k_1$ (KeepOK) & $k_2$ (BackoffOK) & $k_3$ (UpErr) & Net ($k_1-k_2$) \\
\midrule
\multirow{4}{*}{X-Exp-497 (0.479, 137)}
 & Direct\_Delta (Auto-$\Delta$) & 0.527 & 0.573 & $+2.01$ & 0.463 & $-1.61$ & 55 & 63 & 20 & $-8$ ($-7.0$) \\
 & Direct\_AB + $\hat{\pi}_L$ & 0.568 & 0.549 & $-0.20$ & 0.477 & $-0.20$ & 58 & 59 & 20 & $-1$ \\
 & SymDiff + $\hat{\pi}_L$ & 0.675 & \textbf{0.503} & $+2.82$ & 0.523 & $+4.43$ & 68 & 46 & 25 & $+22$ ($+21.3$) \\
 & \textbf{\texttt{AdaDepth} (Ours)} & \textbf{0.710} & \textbf{0.503} & -- & \textbf{0.533} & $\mathbf{+5.43}$ & \textbf{70} & \textbf{43} & 24 & $\mathbf{+27}$ \\
\midrule
\multirow{4}{*}{X-Pilot-200 (0.675, 33)}
 & Direct\_Delta (Auto-$\Delta$) & 0.516 & 0.303 & $+1.00$ & 0.740 & $+6.50$ & 21 & 8 & 4 & $+13$ \\
 & Direct\_AB + $\hat{\pi}_L$ & 0.644 & 0.215 & $+6.00$ & 0.745 & $+7.00$ & \textbf{22} & 8 & 3 & $+14$ \\
 & SymDiff + $\hat{\pi}_L$ & \textbf{0.660} & 0.206 & $+4.00$ & \textbf{0.750} & $\mathbf{+7.50}$ & \textbf{22} & \textbf{7} & 4 & $\mathbf{+15}$ \\
 & \textbf{\texttt{AdaDepth} (Ours)} & 0.652 & \textbf{0.197} & -- & 0.745 & $+7.00$ & \textbf{22} & 8 & 3 & $+14$ \\
\midrule
\multirow{5}{*}{HN-CP1-320 (0.497, 23)}
 & Direct\_Delta (Auto-$\Delta$) & 0.901 & 0.714 & $+0.94$ & 0.509 & $+1.25$ & 11 & 7 & 5 & $+4$ \\
 & Direct\_Delta (Manual-$\Delta$) & \textbf{0.950} & \textbf{0.703} & $+0.63$ & \textbf{0.525} & $\mathbf{+2.81}$ & \textbf{15} & 6 & 3 & $\mathbf{+9}$ ($+8.0$) \\
 & Direct\_AB + $\hat{\pi}_L$ & 0.891 & 0.719 & $-1.87$ & 0.503 & $+0.63$ & 11 & 9 & 3 & $+2$ \\
 & SymDiff + $\hat{\pi}_L$ & 0.853 & 0.772 & $+0.31$ & 0.506 & $+0.94$ & 8 & \textbf{5} & 10 & $+3$ \\
 & \textbf{\texttt{AdaDepth} (Ours)} & 0.878 & 0.734 & -- & 0.503 & $+0.63$ & 9 & 7 & 7 & $+2$ \\
\midrule
\multirow{5}{*}{HN-Reserve-160 (0.575, 10)}
 & Direct\_Delta (Auto-$\Delta$) & 0.873 & 0.712 & $+0.63$ & \textbf{0.588} & $\mathbf{+1.25}$ & \textbf{6} & \textbf{4} & 0 & $\mathbf{+2}$ \\
 & Direct\_Delta (Manual-$\Delta$) & \textbf{0.879} & \textbf{0.706} & $+0.63$ & \textbf{0.588} & $\mathbf{+1.25}$ & \textbf{6} & \textbf{4} & 0 & $\mathbf{+2}$ \\
 & Direct\_AB + $\hat{\pi}_L$ & 0.794 & 0.749 & $-2.50$ & 0.569 & $-0.62$ & 4 & 5 & 2 & $-1$ ($-1.5$) \\
 & SymDiff + $\hat{\pi}_L$ & 0.682 & 0.795 & $+0.00$ & 0.569 & $-0.62$ & 3 & \textbf{4} & 3 & $-1$ \\
 & \textbf{\texttt{AdaDepth} (Ours)} & 0.658 & 0.775 & -- & 0.563 & $-1.25$ & 3 & 5 & 2 & $-2$ \\
\midrule
\multirow{4}{*}{HN-All-480 (0.523, 33)}
 & Direct\_Delta (Auto-$\Delta$) & -- & -- & -- & 0.535 & $+1.25$ & 17 & 11 & 5 & $+6$ \\
 & Direct\_Delta (Manual-$\Delta$) & -- & -- & -- & \textbf{0.546} & $\mathbf{+2.29}$ & \textbf{21} & \textbf{10} & 3 & $\mathbf{+11}$ ($+10.0$) \\
 & Direct\_AB + $\hat{\pi}_L$ & -- & -- & -- & 0.525 & $+0.21$ & 15 & 14 & 5 & $+1$ ($+0.5$) \\
 & \textbf{\texttt{AdaDepth} (Ours)} & -- & -- & -- & 0.523 & $+0.00$ & 12 & 12 & 9 & $0$ \\
\bottomrule
\end{tabular}%
}
\end{table}

\subsection{Cross-platform Pareto, Risk--Coverage, and Accuracy--Specificity analysis}
\begin{figure}[t]
\centering
\includegraphics[width=\textwidth]{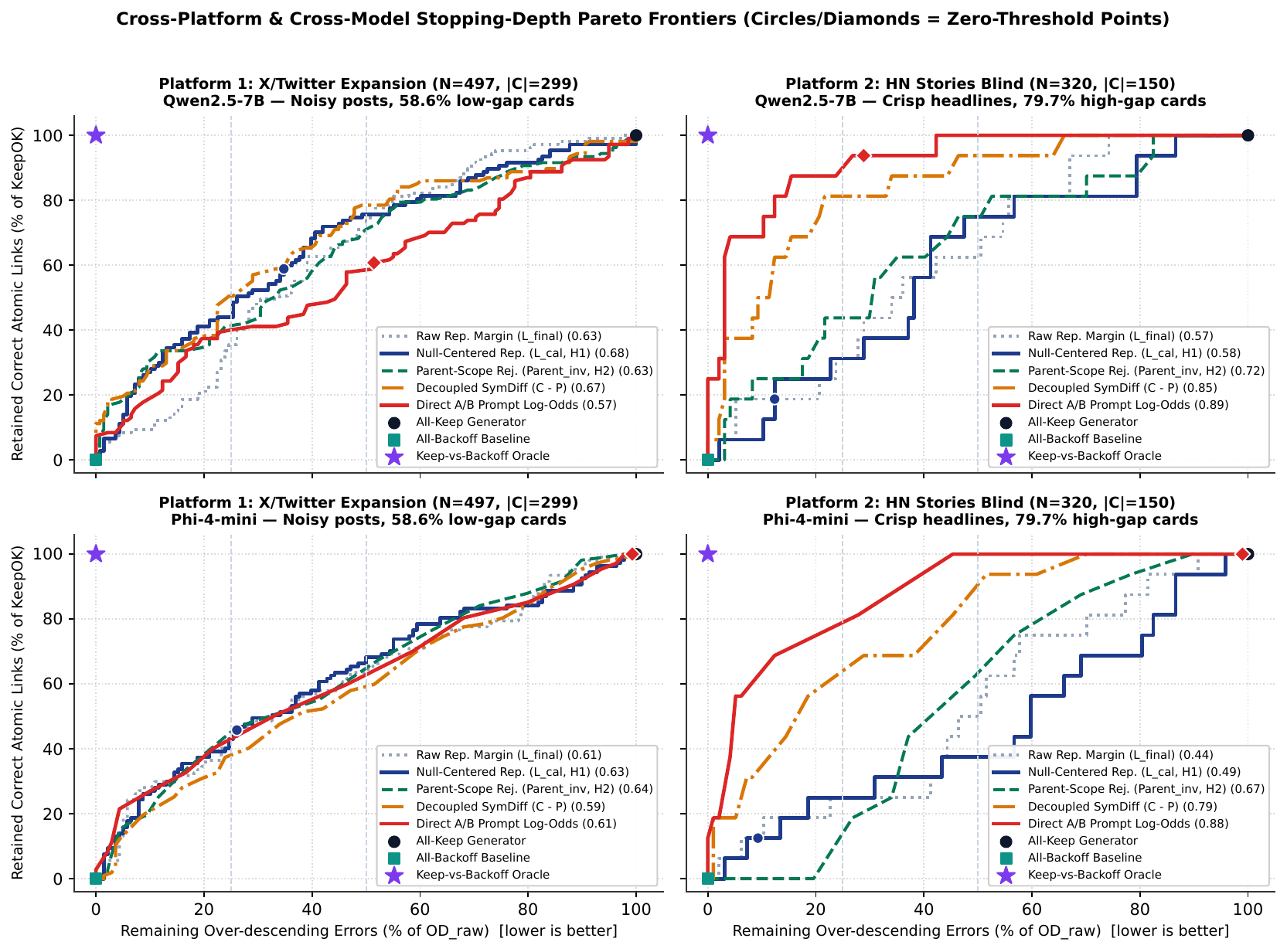}
\caption{Cross-platform and cross-model Pareto frontiers (2$\times$2) for elementary stopping-depth signals. Circles and diamonds mark zero-threshold ($\tau=0$) operating points for $L_{\mathrm{cal}}$ and \texttt{Direct\_AB}. On noisy low-gap X/Twitter posts (left), $L_{\mathrm{cal}}$ and \texttt{SymDiff} lead. On crisp high-gap HN Stories headlines (right), \texttt{Direct\_AB} and \texttt{SymDiff} dominate ranking (0.79--0.89~AUC), yet \texttt{Phi-4-mini}'s uncalibrated \texttt{Direct\_AB} diamond collapses onto the All-Keep corner (96/97 OD), motivating \texttt{AdaDepth}'s null-anchored quantile gate.}
\label{fig:pareto_2x2}
\Description{Two-by-two grid of Atomic retention versus over-descending Pareto frontiers for Qwen2.5-7B and Phi-4-mini across X-Expansion and HN-Stories.}
\end{figure}

\begin{figure}[t]
\centering
\includegraphics[width=\textwidth]{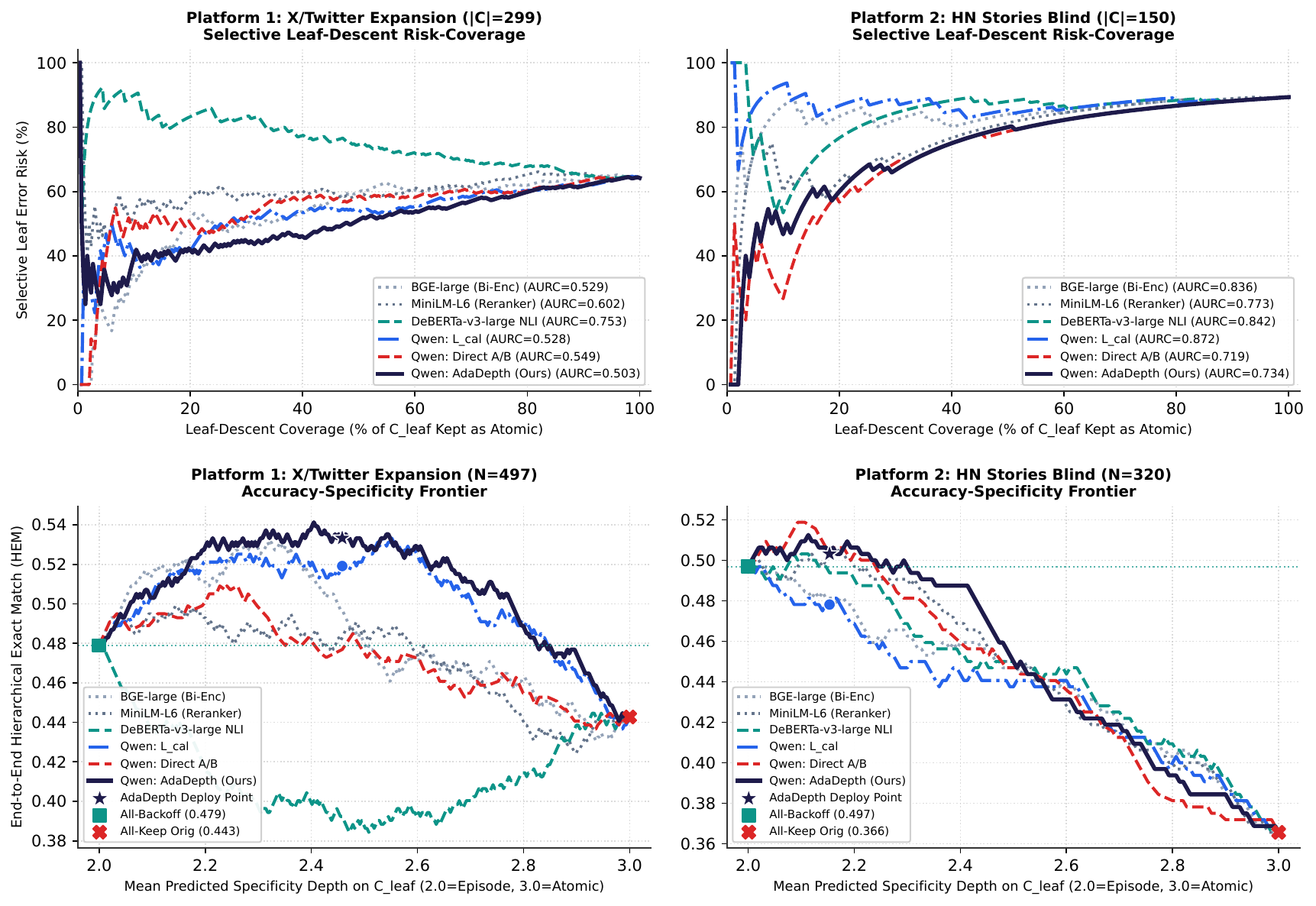}
\caption{IR decision-theoretic evaluation across Platform~1 (X/Twitter Expansion, left) and Platform~2 (HN Stories Blind, right), comparing \texttt{BGE-large}, \texttt{MiniLM-L6}, \texttt{DeBERTa-v3-large}, \texttt{Qwen} $L_{\mathrm{cal}}$, \texttt{Qwen} \texttt{Direct\_AB}, and \texttt{Qwen} \texttt{AdaDepth}. \textbf{Top row:}~Selective leaf-descent Risk--Coverage curves (lower AURC is better), where \texttt{AdaDepth} attains the lowest selective risk on X/Twitter (0.503) and tracks the optimal low-coverage envelope on HN Stories (0.734). \textbf{Bottom row:}~Accuracy--Specificity frontiers sweeping mean predicted ontology depth $\bar{D}\in[2.0, 3.0]$; stars mark \texttt{AdaDepth}'s zero-label operating point (landing cleanly near the frontier apex on both platforms), squares mark All-Backoff ($\bar{D}=2.0$), and crosses mark All-Keep ($\bar{D}=3.0$).}
\label{fig:rc_as}
\Description{Two-by-two grid comparing BGE-large, MiniLM-L6, DeBERTa-v3-large, Qwen L_cal, Qwen Direct A/B, and Qwen AdaDepth across X-Expansion (left) and HN-Stories (right). Top row plots Selective Leaf-Descent Risk-Coverage curves with AURC values; bottom row plots Accuracy-Specificity frontiers with star markers for AdaDepth deploy points, square markers for All-Backoff, and cross markers for All-Keep.}
\end{figure}

\Cref{fig:pareto_2x2} plots the $2\times 2$ cross-platform Atomic retention vs.\ over-descending Pareto frontiers for elementary signals, illustrating both the sharp platform regime inversion and \texttt{Phi-4-mini}'s zero-threshold logit collapse on HN Stories.
\Cref{fig:rc_as} deepens this analysis through two standard IR decision curves, \textbf{Risk--Coverage} (top row) and \textbf{Accuracy--Specificity} (bottom row), comparing \texttt{AdaDepth} directly against \texttt{DeBERTa-v3-large}, \texttt{BGE-large}, \texttt{MiniLM-L6}, \texttt{MSP}, and \texttt{DARTS}:
\begin{itemize}[leftmargin=*,nosep]
  \item \textbf{Risk--Coverage (AURC / E-AURC):} On X-Expansion (\Cref{fig:rc_as}, top-left), \texttt{AdaDepth} traces the lowest selective leaf-risk envelope across nearly the entire coverage range ($\mathrm{AURC}=0.503$, $\mathrm{E\text{-}AURC}=0.228$), outperforming \texttt{BGE-large} (0.529), \texttt{DARTS}/\texttt{Direct\_AB} (0.549), \texttt{MiniLM-L6} (0.602), \texttt{MSP} (0.648, \Cref{tab:tois_master}), and \texttt{DeBERTa-v3-large} (0.753, which rises \emph{above} random ordering). On blind HN Stories (\Cref{fig:rc_as}, top-right), \texttt{AdaDepth} hugs the optimal risk envelope at low-to-moderate coverage ($\mathrm{AURC}=0.734$, $\mathrm{E\text{-}AURC}=0.077$), beating \texttt{MiniLM-L6} (0.773), \texttt{BGE-large} (0.837), \texttt{DeBERTa} (0.842), and \texttt{MSP} (0.878).
  \item \textbf{Accuracy--Specificity Frontier:} Sweeping mean output depth $\bar{D}\in[2.0, 3.0]$ (\Cref{fig:rc_as}, bottom row) demonstrates that \texttt{AdaDepth}'s star operating point sits cleanly near the peak of the concave accuracy--specificity frontier on \emph{both} platforms ($\bar{D}=2.46$ on X-Expansion; $\bar{D}=2.15$ on HN Stories). By contrast, \texttt{BGE-large} and \texttt{MSP} cluster near the bottom-right over-descending cliff ($\bar{D}=2.82$--$2.96$, \Cref{tab:tois_master}), trading away $4$--$11$~pp of exact-match accuracy for hallucinated specificity.
\end{itemize}

\subsection{Paired statistical inference and computational efficiency}
\Cref{tab:stat_tests} compiles 10,000-resample paired bootstrap 95\% confidence intervals and exact two-sided McNemar $p$-values across both platforms and models.
Two key statistical facts emerge: (i)~\textbf{H1 ($L_{\mathrm{cal}}$ over $L_{\mathrm{raw}}$)} improves all 4/4 model--platform cells, with significant AUC gain on X-Expansion Qwen ($+0.0488$, CI $[+0.0070,+0.0932]$) and significant deployable $\mathrm{HEM}(0)$ gains on HN Stories for both Qwen ($+2.50$~pp, $p=0.0386$) and Phi-4 ($+6.88$~pp, $p=0.0003$), while strictly verifying $|\mathrm{InPairAUC}(L_{\mathrm{cal}})-\mathrm{InPairAUC}(L_{\mathrm{raw}})| < 10^{-12}$ on every run; (ii)~the \textbf{cross-platform regime inversion} between \texttt{Direct\_AB} and $L_{\mathrm{cal}}$ has 95\% CIs excluding zero in \emph{opposite directions} on Qwen2.5-7B ($[-0.2129, -0.0105]$ on X-Expansion vs.\ $[+0.1325, +0.4817]$ on HN Stories); and (iii)~\textbf{\texttt{AdaDepth}} yields large, statistically significant gains over the base generator across every platform and model ($+7.04$ to $+13.75$~pp, all $p \le 0.0043$).

\begin{table}[t]
\centering
\caption{Paired statistical inference across both platforms and models (10,000 paired bootstrap resamples, seed \texttt{20261128}; exact two-sided McNemar $p$-values). Bold confidence intervals exclude zero.}
\label{tab:stat_tests}
\footnotesize
\setlength{\tabcolsep}{3.0pt}
\resizebox{\textwidth}{!}{%
\begin{tabular}{llccccc}
\toprule
Platform & Model & H1 $\Delta\mathrm{HEM}(0)$ [$L_{\mathrm{cal}}-L_{\mathrm{raw}}$] ($p$) & H1 $\Delta\mathrm{AUC}$ [$L_{\mathrm{cal}}-L_{\mathrm{raw}}$] & Regime $\Delta\mathrm{AUC}$ [\texttt{Dir}$-L_{\mathrm{cal}}$] & \texttt{AdaDepth} vs.\ Orig $\Delta$HEM ($p$) \\
\midrule
X-Exp (497) & Qwen2.5-7B & $+0.0000$ $[-0.030, +0.028]$ ($1.000$) & $\mathbf{+0.0488\ [+0.007, +0.093]}$ & $\mathbf{-0.1135\ [-0.213, -0.011]}$ & $\mathbf{+0.0905}$ ($4.5\times 10^{-5}$) \\
X-Exp (497) & Phi-4-mini & $+0.0262$ $[-0.014, +0.064]$ ($0.232$) & $+0.0180$ $[-0.056, +0.092]$ & $-0.0225$ $[-0.121, +0.078]$ & $\mathbf{+0.0704}$ ($0.0043$) \\
HN (320) & Qwen2.5-7B & $\mathbf{+0.0250\ [+0.006, +0.047]}$ (\textbf{0.039}) & $+0.0108$ $[-0.105, +0.132]$ & $\mathbf{+0.3141\ [+0.133, +0.482]}$ & $\mathbf{+0.1375}$ ($2.4\times 10^{-9}$) \\
HN (320) & Phi-4-mini & $\mathbf{+0.0688\ [+0.034, +0.106]}$ (\textbf{0.0003}) & $+0.0442$ $[-0.137, +0.236]$ & $\mathbf{+0.3912\ [+0.252, +0.534]}$ & $\mathbf{+0.1344}$ ($1.8\times 10^{-8}$) \\
\bottomrule
\end{tabular}%
}
\end{table}

\textbf{System efficiency and deployment cost.}
\Cref{tab:efficiency} logs exact GPU memory footprints and wall-clock throughput measured on a single NVIDIA RTX~PRO~4500 GPU.
Because \texttt{AdaDepth} performs strictly \emph{single-forward-pass} embedding and first-token logit extraction (zero autoregressive token generation) with offline Event Card caching, the compact 3.8B \texttt{Phi-4-mini} backbone runs inside 7.3~GB VRAM at 9.7~items/s composite throughput (103~ms/candidate, or 24~ms/candidate for $L_{\mathrm{cal}}$ alone), while \texttt{Qwen2.5-7B} runs inside 14.5~GB VRAM at 5.7~items/s (176~ms/candidate), making \texttt{AdaDepth} practical for real-time post-proposal filtering in production IR pipelines.

\begin{table}[t]
\centering
\caption{Measured computational cost on a single NVIDIA RTX~PRO~4500 GPU (bf16/fp16 inference, zero autoregressive decoding). Event Card embeddings and 3-null baselines are cached once offline.}
\label{tab:efficiency}
\footnotesize
\setlength{\tabcolsep}{4.8pt}
\begin{tabular}{llcccc}
\toprule
Model Family & Backbone / Signal & Params & Peak VRAM (MB) & Forward Passes / Cand. & Throughput (cand./s) \\
\midrule
Dense IR & BGE-large-en-v1.5 & 335M & 639.6 & 1 (post enc.) & 285.6 \\
Cross-Enc. IR & MiniLM-L6-v2 & 22.7M & 43.9 & 2 ($P$ + $C$) & 187.4 \\
NLI Classifier & DeBERTa-v3-large & 435M & 829.8 & 2 ($P$ + $C$) & 66.8 \\
\midrule
\multirow{4}{*}{Phi-4-mini}
 & $L_{\mathrm{cal}}$ (Path B only) & 3.8B & 7,316.6 & 1 (post enc.) & 40.9--48.1 \\
 & Direct\_AB (Path C) & 3.8B & 7,316.6 & 1 (1st-token) & 26.4 \\
 & SymDiff (Path C) & 3.8B & 7,316.6 & 2 (Yes/No $P,C$) & 23.2 \\
 & \textbf{Full \texttt{AdaDepth}} & 3.8B & 7,316.6 & 4 (cached cards) & \textbf{9.7} (103~ms) \\
\midrule
\multirow{4}{*}{Qwen2.5-7B}
 & $L_{\mathrm{cal}}$ (Path B only) & 7.6B & 14,529.1 & 1 (post enc.) & 30.0 \\
 & Direct\_AB (Path C) & 7.6B & 14,529.1 & 1 (1st-token) & 15.3--15.8 \\
 & SymDiff (Path C) & 7.6B & 14,529.1 & 2 (Yes/No $P,C$) & 13.4 \\
 & \textbf{Full \texttt{AdaDepth}} & 7.6B & 14,529.1 & 4 (cached cards) & \textbf{5.7} (176~ms) \\
\bottomrule
\end{tabular}
\end{table}

\section{Mechanistic Analysis: Causal Falsifications, Granularity-Gap Unification, and Faithfulness}\label{sec:mech}
Why does \texttt{AdaDepth}'s dual-path design work, and do its signals genuinely track human-annotated incremental evidence rather than superficial heuristics? We answer this via three controlled causal interventions.

\subsection{Three causal falsifications across models and platforms}
\textbf{Falsification 1: Free-form LLM self-audit (\texttt{T2}) collapses.}
Prompting \texttt{Qwen2.5-7B} to extract verbatim parent and child evidence quotes before scoring A vs.\ B drops ranking to near-random (\textbf{0.4948~AUC} on X-Expansion). Qualitative audit reveals sycophantic conservatism: the model fabricates parent evidence on 291/299 items while returning \texttt{NONE} for child evidence on 251/299 items.

\begin{figure}[t]
\centering
\includegraphics[width=\textwidth]{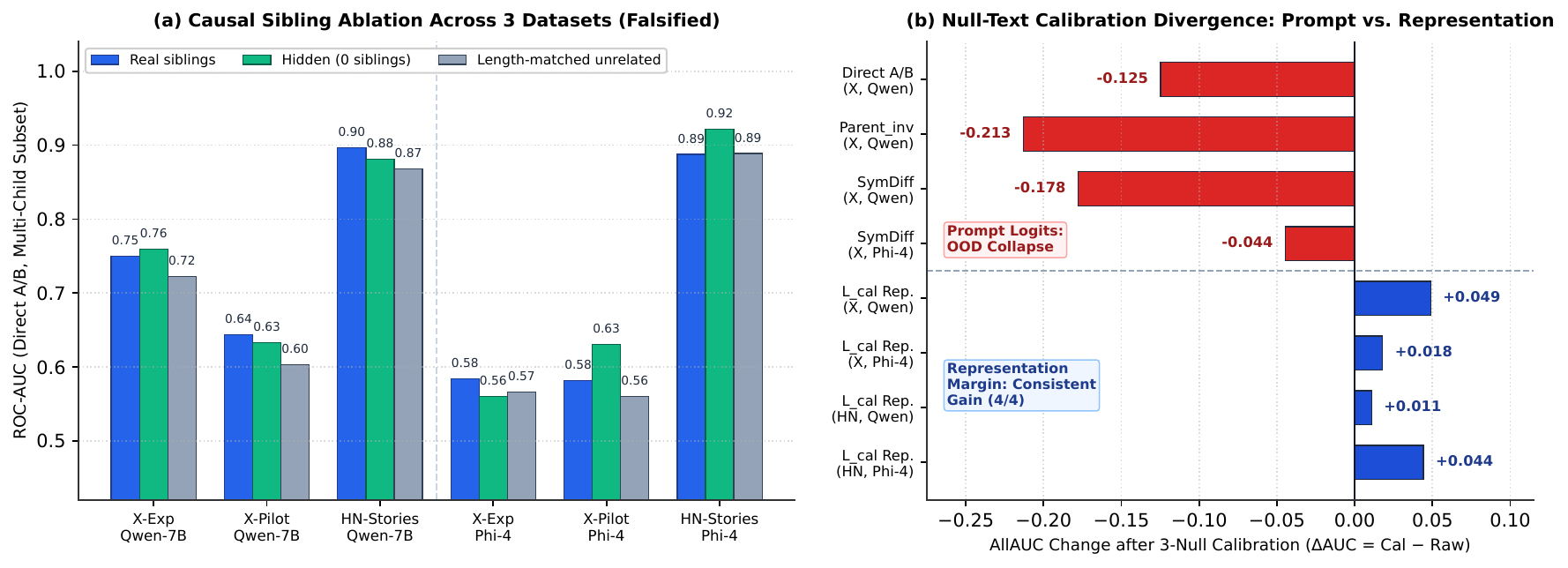}
\caption{Controlled causal falsifications across platforms and models. \textbf{(a)}~3-dataset causal sibling ablation (\texttt{real} vs.\ \texttt{hidden} vs.\ length-matched \texttt{unrelated}) proves removing sibling cards never collapses stopping-depth discrimination. \textbf{(b)}~3-null contextual calibration triggers OOD collapse on prompt logits (red) while consistently improving hidden-state margins $L_{\mathrm{cal}}$ across all four model--platform cells (blue).}
\label{fig:causal_cal}
\Description{Two-panel chart showing the 3-dataset causal sibling ablation on the left and the divergence of 3-null contextual calibration between prompt logits and hidden-state margins on the right.}
\end{figure}

\textbf{Falsification 2: Prompt-visible sibling cards do not cause multi-child discrimination.}
Because \texttt{Direct\_AB} scored higher on multi-child X-Expansion items (0.750~AUC) than single-child items (0.514~AUC), a natural hypothesis was that listing sibling Atomic cards inside the prompt sharpens contrast.
We falsified this directly across \emph{all three datasets} (64 X-Expansion, 35 X-Pilot, and 135 HN-Stories multi-child candidates; 6 dataset--model cells in \Cref{fig:causal_cal}a) by comparing \texttt{real} sibling lists against \texttt{hidden} (zero siblings) and length-matched \texttt{unrelated} cards from other Storylines.
Hiding sibling cards \emph{never} collapses discrimination and actually improves AUC in half of all cells: \texttt{Qwen2.5-7B} X-Exp (\texttt{real} 0.7500 vs.\ \texttt{hidden} 0.7596), \texttt{Phi-4-mini} X-Pilot (0.5815 vs.\ 0.6304), and \texttt{Phi-4-mini} HN-Stories (\texttt{real} 0.8875 vs.\ \textbf{\texttt{hidden} 0.9218}, $+3.43$~pp; \texttt{Qwen2.5-7B} HN: 0.8969 vs.\ 0.8814).
Prompt-visible sibling contrast is therefore \emph{not} the causal driver.

\textbf{Falsification 3: Content-free 3-null calibration diverges sharply between logits and representations.}
As shown in \Cref{fig:causal_cal}b, subtracting 3-null baselines from instruction-tuned prompt logits severely degrades ranking on \texttt{Qwen2.5-7B} X-Expansion (\texttt{Direct\_AB}: $0.568\to 0.443$, $-0.125$; \texttt{Parent\_inv}: $0.634\to 0.421$, $-0.213$; \texttt{SymDiff}: $0.675\to 0.497$, $-0.178$) and similarly harms \texttt{BART-large-MNLI} ($-0.112$ on X-Exp, $-0.189$ on HN-320).
Inspection reveals that instruction-tuned heads treat empty/\texttt{"N/A"} prompts as out-of-distribution refusal triggers rather than neutral class priors, whereas final-layer mean-pooled embeddings respond smoothly to static card-boilerplate bias, enabling $L_{\mathrm{cal}}$ to improve all 4/4 model--platform pairs.

\subsection{Granularity-Gap unification across platforms}
If prompt-visible siblings do not explain why multi-child items favor prompt log-odds, what does?
Stratifying X-Expansion by card-intrinsic Granularity Gap (\Cref{tab:gap_strat}, \Cref{fig:gap_faith}a) uncovers the unifying variable: mean parent--child hidden-state cosine similarity $\mathrm{sim}(c_e,c_a)$ decreases strictly monotonically from \texttt{low} to \texttt{high} gap on both models (\texttt{Qwen}: $0.943\to 0.936\to 0.913$; \texttt{Phi-4}: $0.985\to 0.980\to 0.977$).
When the gap is \texttt{low} ($n_{\mathrm{dec}}=136$, 58.6\% of X-Expansion), $L_{\mathrm{cal}}$ leads (0.760 Qwen, 0.741 Phi-4) while \texttt{Direct\_AB} stalls (0.537 Qwen); when the gap is \texttt{high} ($n_{\mathrm{dec}}=65$), \texttt{SymDiff} (0.728 Qwen) and \texttt{Direct\_AB} (0.678 Qwen, 0.696 Phi-4) overtake it.
This within-platform transition reproduces the shift observed on high-gap HN Stories (79.7\% \texttt{high}-gap decisive items, where \texttt{Direct\_AB} reaches 0.925/0.890~AUC), explaining why \texttt{AdaDepth}'s structural gap gate $w_{\mathrm{gap}}(e,a)$ outperforms fixed-weight mixing across regimes.

\begin{table}[t]
\centering
\caption{Granularity-Gap ($\Delta(e,a)$) stratification unifying Platform~1 (X-Expansion) and Platform~2 (HN-Stories). As parent--child overlap ($\mathrm{sim}_{PC}$) falls from \texttt{low} to \texttt{high} gap, prompt log-odds overtake representation margins $L_{\mathrm{cal}}$ both within X/Twitter and across to HN Stories.}
\label{tab:gap_strat}
\footnotesize
\setlength{\tabcolsep}{4.0pt}
\resizebox{\textwidth}{!}{%
\begin{tabular}{llccccccc}
\toprule
Platform & Stratum & $n_{\mathrm{dec}}$ (pos) & $\mathrm{sim}_{PC}$ (Qwen) & $L_{\mathrm{cal}}$ (Q / P) & \texttt{Parent\_inv} (Q / P) & \texttt{Child} (Q / P) & \texttt{SymDiff} (Q / P) & \texttt{Direct\_AB} (Q / P) \\
\midrule
X-Exp & \texttt{gap=low} & 136 (41) & 0.9429 & \textbf{0.760} / \textbf{0.741} & 0.652 / 0.650 & 0.515 / 0.560 & 0.645 / 0.627 & 0.537 / 0.637 \\
X-Exp & \texttt{gap=medium} & 31 (15) & 0.9357 & 0.363 / 0.696 & 0.500 / 0.573 & 0.602 / 0.671 & 0.583 / \textbf{0.733} & \textbf{0.708} / 0.523 \\
X-Exp & \texttt{gap=high} & 65 (51) & 0.9125 & 0.598 / 0.622 & 0.576 / 0.663 & 0.698 / 0.575 & \textbf{0.728} / 0.661 & 0.678 / \textbf{0.696} \\
\midrule
HN-320 & \texttt{gap=high} & 59 (12) & -- & 0.582 / 0.493 & 0.781 / 0.680 & 0.646 / 0.689 & 0.920 / 0.813 & \textbf{0.925} / \textbf{0.890} \\
\bottomrule
\end{tabular}%
}
\end{table}

\subsection{Counterfactual evidence-span faithfulness}
\begin{figure}[t]
\centering
\includegraphics[width=\textwidth]{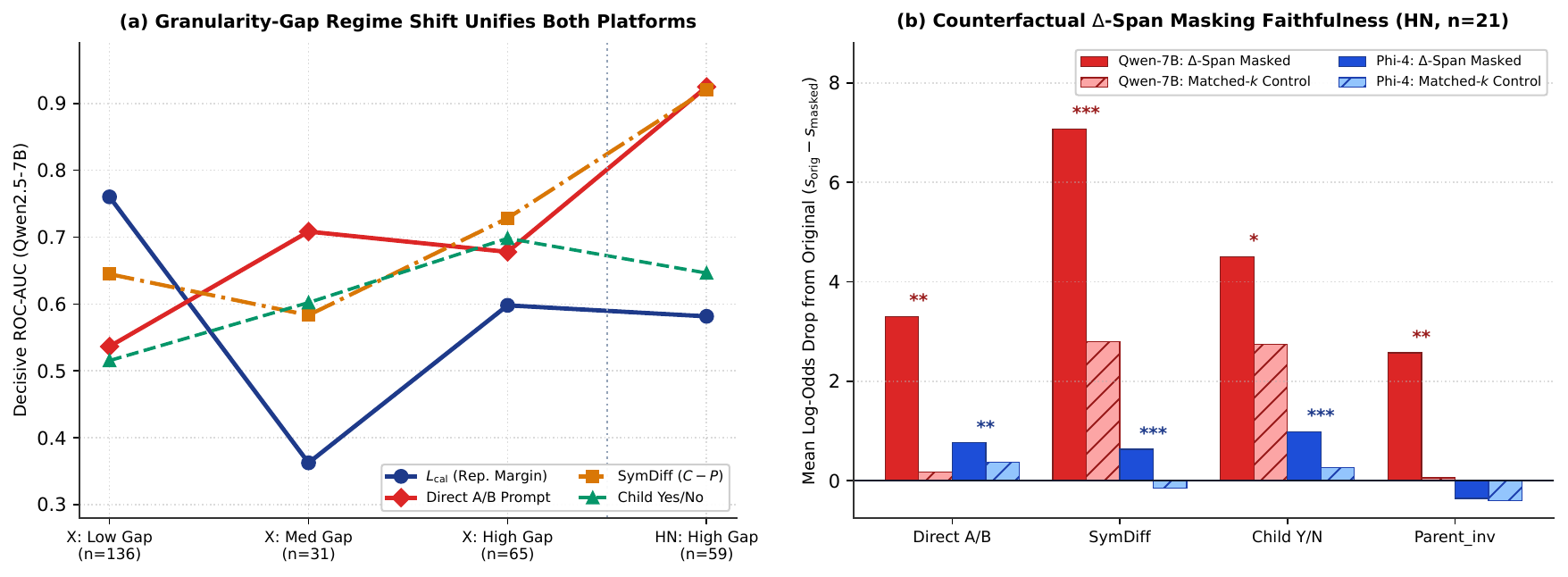}
\caption{Mechanistic unification and counterfactual faithfulness. \textbf{(a)}~Card-level Granularity Gap reproduces the shift between $L_{\mathrm{cal}}$ and prompt log-odds \emph{within} X/Twitter and extends continuously to high-gap HN Stories. \textbf{(b)}~Matched-$k$ counterfactual masking confirms that prompt log-odds drop sharply when incremental $\Delta$-tokens inside $S_{\mathrm{inc}}(p)$ are masked relative to length-matched controls, whereas mean-pooled $L_{\mathrm{cal}}$ is insensitive to surgical token edits.}
\label{fig:gap_faith}
\Description{Two-panel mechanistic chart showing Granularity-Gap unification across X-Expansion and HN-Stories on the left and matched-k counterfactual span-masking faithfulness drops on the right.}
\end{figure}

To test whether stopping-depth signals causally respond to human-annotated incremental evidence spans $S_{\mathrm{inc}}(p)$ rather than shared entity cues, we run matched-$k$ counterfactual token masking on all $n=21$ Gold \texttt{atomic\_exact} HN headlines (\Cref{tab:cf_faithfulness}, \Cref{fig:gap_faith}b).
For each headline, we deterministically partition content tokens into $\Delta$-tokens (inside $S_{\mathrm{inc}}(p)$ and absent from parent card $c_e$) versus non-$\Delta$ anchor tokens, masking the \emph{exact same count $k$} in paired counterfactuals (\texttt{delta\_masked} vs.\ \texttt{ctrl\_masked}).

\begin{table}[t]
\centering
\caption{Counterfactual evidence-span masking faithfulness on HN Stories Gold \texttt{atomic\_exact} headlines ($n=21$; matched-$k$ token masking comparing \texttt{delta\_masked} vs.\ \texttt{ctrl\_masked}). Positive $\mathrm{Net} = s_{\mathrm{ctrl}} - s_{\mathrm{delta}}$ confirms causal dependence on incremental $\Delta(e,a)$ tokens.}
\label{tab:cf_faithfulness}
\footnotesize
\setlength{\tabcolsep}{3.6pt}
\resizebox{\textwidth}{!}{%
\begin{tabular}{llccccccc}
\toprule
Model & Signal & Orig & Ctrl-Masked & $\Delta$-Masked & Drop ($\Delta$) & Drop (Ctrl) & Net (Ctrl $-\Delta$) [95\% CI] & Wilcoxon $p$ \\
\midrule
\multirow{5}{*}{Qwen2.5-7B}
 & Direct\_AB & $+7.321$ & $+7.143$ & $+4.024$ & $+3.298$ & $+0.179$ & $\mathbf{+3.119\ [+1.351, +4.923]}$ & \textbf{0.0024} \\
 & SymDiff ($C-P$) & $-13.357$ & $-16.155$ & $-20.435$ & $+7.077$ & $+2.798$ & $\mathbf{+4.280\ [+2.375, +5.970]}$ & \textbf{0.00035} \\
 & Child Yes/No & $-7.321$ & $-10.060$ & $-11.827$ & $+4.506$ & $+2.738$ & $\mathbf{+1.768\ [+0.214, +3.345]}$ & \textbf{0.0290} \\
 & Parent\_inv & $-6.036$ & $-6.095$ & $-8.607$ & $+2.571$ & $+0.060$ & $\mathbf{+2.512\ [+1.309, +3.714]}$ & \textbf{0.0012} \\
 & $L_{\mathrm{cal}}$ (L28 mean) & $-0.032$ & $-0.040$ & $-0.036$ & $+0.004$ & $+0.008$ & $-0.003\ [-0.017, +0.010]$ & 0.6460 \\
\midrule
\multirow{5}{*}{Phi-4-mini}
 & Direct\_AB & $+2.607$ & $+2.238$ & $+1.845$ & $+0.762$ & $+0.369$ & $\mathbf{+0.393\ [+0.143, +0.702]}$ & \textbf{0.0033} \\
 & SymDiff ($C-P$) & $-6.738$ & $-6.595$ & $-7.369$ & $+0.631$ & $-0.143$ & $\mathbf{+0.774\ [+0.369, +1.226]}$ & \textbf{0.00086} \\
 & Child Yes/No & $-3.000$ & $-3.262$ & $-3.988$ & $+0.988$ & $+0.262$ & $\mathbf{+0.726\ [+0.369, +1.095]}$ & \textbf{0.00093} \\
 & Parent\_inv & $-3.738$ & $-3.333$ & $-3.381$ & $-0.357$ & $-0.405$ & $+0.048\ [-0.155, +0.238]$ & 0.3685 \\
 & $L_{\mathrm{cal}}$ (L32 mean) & $-0.025$ & $-0.030$ & $-0.031$ & $+0.005$ & $+0.004$ & $+0.001\ [-0.003, +0.005]$ & 0.5942 \\
\bottomrule
\end{tabular}%
}
\end{table}

The results establish a clean mechanistic dissociation across both architectures: prompt log-odds causally track $\Delta$-tokens (\texttt{Qwen} \texttt{SymDiff} $\mathrm{Net}=+4.280$, $p=0.00035$; \texttt{Direct\_AB} $\mathrm{Net}=+3.119$, $p=0.0024$; \texttt{Phi-4} \texttt{SymDiff} $\mathrm{Net}=+0.774$, $p=0.00086$; \texttt{Direct\_AB} $\mathrm{Net}=+0.393$, $p=0.0033$, replicating identically on the $n=16$ frozen \texttt{KeepOK} subset and full-span masking over the $n=10$ partial-span headlines, e.g., \texttt{Qwen} \texttt{Direct\_AB} $\mathrm{Net}=+8.81$, $p=0.0029$).
In contrast, mean-pooled $L_{\mathrm{cal}}$ is unaffected by surgical $\Delta$-token masking (\texttt{Qwen} $p=0.646$; \texttt{Phi-4} $p=0.594$), directly explaining why representation margins smooth out noisy social posts yet require \texttt{AdaDepth}'s gap-gated fusion with prompt log-odds whenever a single incremental verb or ruling dictates stopping depth.

\section{Discussion, Limitations, and Conclusion}\label{sec:discussion}

\subsection{Design implications for hierarchical event linking}
Our findings yield three concrete principles for designing production event-linking and hierarchical IR systems:
\begin{enumerate}[leftmargin=*,nosep]
  \item \textbf{Decouple relevance retrieval from evidential stopping depth.} Dense bi-encoders, rerankers, and NLI models remain valuable for candidate generation, yet deploying them as stopping-depth gates induces severe over-descending ($4$--$11$~pp below All-Backoff). Systems should pair open-set scope gating and deterministic validation $V_{\mathcal{H}}$ with a dedicated stopping-depth calibrator.
  \item \textbf{Anchor deployable thresholds to unlabeled representation priors.} Prompt log-odds rank crisp incremental assertions accurately (0.85--0.91~AUC) but exhibit unpredictable cross-model offset drift at $\tau=0$. Anchoring the fused score to the unlabeled 3-null prevalence rate $\hat{\pi}_L$ eliminates drift without a single labeled target example.
  \item \textbf{Author Event Cards around $\Delta(e,a)$ from day one.} Explicitly specifying Incremental Assertions and disqualifying parent framings sharpens annotator agreement, enables offline computation of $w_{\mathrm{gap}}(e,a)$, and flags redundant single-child leaves (like \texttt{EP31\_02}) before deployment.
\end{enumerate}

\subsection{Limitations and future directions}
First, our benchmark spans 1,177 English records across nine Storylines on X/Twitter and Hacker News; extending evidence-supported stopping depth to multilingual and multimodal (image, video, quote-thread) streams remains open.
Second, because rare \texttt{atomic\_exact} headlines occur at low base rates on Hacker News ($n=21$ in CP1, $n=12$ in Reserve), the All-Backoff-to-Oracle ceiling is narrow (5.0~pp) and finite-sample variance in $\hat{\pi}_L$ is more visible on small candidate batches; furthermore, \texttt{HN-CP1} and \texttt{HN-Reserve} share the same three-Storyline HN ontology, so prospective validation on an independent unseen Storyline ontology remains future work.
Third, although \texttt{AdaDepth} consumes only standard parent--child card descriptions $(c_e, c_a)$ and our \texttt{Auto-$\Delta$} probe confirms that explicit $\Delta(e,a)$ criteria can be extracted automatically from those descriptions without manual authoring, evaluating our stopping-depth framework directly on legacy public event hierarchies (e.g., WikiEvents or ACE), deeper multi-hop taxonomies (depth $>3$), and streaming distribution-free conformal risk control under online drift are important directions for future work.

\subsection{Conclusion}
We investigated a foundational decision problem in structured information retrieval: \emph{when semantic relevance is insufficient to substantiate fine-grained event linking, how can an information system detect the evidential sufficiency boundary and decide when to stop descending?}
Through a two-platform, nine-Storyline benchmark totaling 1,177 annotated records (640 dual-audited, including a pre-annotation consistency-checked prospective Stage-2 blind split) and evaluation of 15+ modern IR, reranker, NLI, selective-prediction, and hierarchical-hedging baselines, we demonstrated that topical relevance and standard entailment systematically fail at stopping-depth calibration, highlighting the All-Backoff Paradox.
By integrating open-set scope gating, legal path decoding, representation-level 3-null centering, and $\Delta$-faithful prompt log-odds inside our unified \textbf{\texttt{EviLink}} framework (anchored by the zero-parameter \textbf{\texttt{AdaDepth}} stopping-depth calibrator), we achieve a $+2.97$~pp mean gain over All-Backoff across all 8 benchmark cells (7/8 wins; 2/2 on dual-audited \texttt{Reserve-Dual-80}) and the highest mean gain among all $\hat{\pi}_L$-anchored comparators under both inclusive and strict-$K$ tie-breaking, lead all raw-threshold baselines on \texttt{HN-All-480} (while showing that $\hat{\pi}_L$-anchored direct $\Delta(e,a)$ verification excels on crisp high-gap headlines at 0.87--0.95~AUC even with automatically extracted $\widehat{\Delta}_{\mathrm{auto}}$ yet degrades on noisy low-gap social discourse), and advance the Risk--Coverage and Accuracy--Specificity frontiers at real-time single-GPU latency.

\appendix

\section{Layer-Depth, Robustness Envelope, and Stage-2 Freeze Audit}\label{app:layers}\label{app:robustness}
\Cref{tab:app_layers} summarizes the multi-layer depth (50\%, 75\%, 100\%) and pooling (\texttt{mean} vs.\ \texttt{last}) ablation for $L_{\mathrm{cal}}$ across both models and platforms: final-layer mean pooling leads on noisy X/Twitter posts, whereas intermediate layers retain stronger within-pair headline discrimination on HN Stories. \texttt{AdaDepth} fixes final-layer mean pooling uniformly across all splits to preserve zero-parameter domain transfer.

\Cref{tab:app_robust} reports three robustness diagnostics: (i)~\textbf{unlabeled $\hat{\pi}_L$ base-rate tracking error}, showing \texttt{Phi-4-mini} tracks a $6\times$ true Atomic prevalence shift with 2.8~pp MAE (0.67~pp error on both \texttt{X-Exp} and \texttt{HN-CP1}); (ii)~\textbf{structural gap validation and $\alpha$-sensitivity envelope}, verifying that $w_{\mathrm{gap}}(e,a)$ correlates significantly with pre-Gold expert Granularity Gap ratings on \texttt{HN-All-480} (Spearman $\rho=0.403$, $p=4.3\times 10^{-10}$; card-level $\rho=0.581$, $p=0.029$, rising monotonically from \texttt{low} 0.264 to \texttt{medium} 0.785 and \texttt{high} 0.808) and that varying $w_{\mathrm{gap}}(\alpha) = \alpha\,\mathbb{I}[|\mathcal{A}_e|\ge 2] + (1-\alpha)\,d_{\mathrm{lex}}$ across $\alpha\in[0.25,1.0]$ yields a flat envelope ($+2.75$ to $+3.20$~pp mean $\Delta$BO), confirming insensitivity around the fixed $\alpha=0.50$ equal-weight prior; and (iii)~\textbf{transductive batch-size sensitivity}, partitioning \texttt{X-Exp-497} and \texttt{HN-All-480} into 100 random disjoint batches of size $B\in\{16,32,64,128,\text{Full}\}$, showing \texttt{AdaDepth} outperforms All-Backoff on average across every batch size down to $B=16$ ($\mathrm{std} \le 0.007$) and stabilizes for $B \ge 32$.
\Cref{tab:app_freeze}\label{app:freeze} lists the core Stage-2 Blind \texttt{Reserve-160} prediction artifacts frozen prior to dual human annotation and verified for exact file consistency before evaluation (complete integrity checksums are included in the reproducibility package).

\begin{table}[h]
\centering
\caption{Layer-depth and pooling ablation for 3-null calibrated representation margins ($L_{\mathrm{cal}}$).}
\label{tab:app_layers}
\footnotesize
\setlength{\tabcolsep}{5.0pt}
\begin{tabular}{llcccc}
\toprule
Platform & Model & Layer & Pool & AllAUC & HEM(0) \\
\midrule
X-Exp & Qwen-7B & L14 (50\%) & mean / last & 0.596 / 0.507 & 0.455 / 0.447 \\
X-Exp & Qwen-7B & L21 (75\%) & mean / last & 0.557 / 0.533 & 0.481 / 0.451 \\
X-Exp & Qwen-7B & L28 (100\%) & mean / last & \textbf{0.681} / 0.484 & \textbf{0.519} / 0.451 \\
X-Exp & Phi-4 & L16 (50\%) & mean / last & 0.421 / 0.488 & 0.479 / 0.437 \\
X-Exp & Phi-4 & L24 (75\%) & mean / last & 0.415 / 0.474 & 0.479 / 0.453 \\
X-Exp & Phi-4 & L32 (100\%) & mean / last & \textbf{0.628} / 0.473 & \textbf{0.517} / 0.453 \\
\midrule
HN-320 & Qwen-7B & L14 (50\%) & mean / last & 0.566 / \textbf{0.721} & 0.366 / 0.475 \\
HN-320 & Qwen-7B & L21 (75\%) & mean / last & 0.571 / 0.609 & 0.372 / 0.472 \\
HN-320 & Qwen-7B & L28 (100\%) & mean / last & 0.577 / 0.454 & \textbf{0.478} / 0.409 \\
HN-320 & Phi-4 & L16 (50\%) & mean / last & 0.675 / 0.628 & \textbf{0.497} / 0.441 \\
HN-320 & Phi-4 & L24 (75\%) & mean / last & \textbf{0.711} / 0.546 & \textbf{0.497} / 0.475 \\
HN-320 & Phi-4 & L32 (100\%) & mean / last & 0.486 / 0.681 & 0.478 / 0.441 \\
\bottomrule
\end{tabular}
\end{table}

\begin{table}[h]
\centering
\caption{Robustness envelope: (a) Unlabeled $\hat{\pi}_L$ base-rate tracking error vs.\ true leaf rate $\pi^* = \mathrm{KeepOK}/|\mathcal{C}_{\mathrm{leaf}}|$; (b) sensitivity of \texttt{AdaDepth} 8-cell Mean $\Delta$BO (pp) across $\alpha\in\{0, 0.25, 0.50, 0.75, 1.0\}$ in $w_{\mathrm{gap}}(\alpha)=\alpha\,\mathbb{I}[|\mathcal{A}_e|\ge 2]+(1-\alpha)\,d_{\mathrm{lex}}$; (c) transductive batch-size sensitivity (mean $\pm$ std HEM across 100 random disjoint batch partitions of size $B$).}
\label{tab:app_robust}
\footnotesize
\setlength{\tabcolsep}{4.4pt}
\begin{tabular}{lccccc}
\toprule
\multicolumn{6}{c}{\textbf{(a) Unlabeled $\hat{\pi}_L$ Base-Rate Tracking Across $6\times$ Shift}} \\
\midrule
Dataset ($|\mathcal{C}_{\mathrm{leaf}}|$) & True $\pi^*$ & Qwen $\hat{\pi}_L$ & Qwen $|\mathrm{Err}|$ & Phi-4 $\hat{\pi}_L$ & Phi-4 $|\mathrm{Err}|$ \\
\midrule
X-Pilot (35) & 0.657 & 0.943 & 0.286 & 0.743 & 0.086 \\
X-Expansion (299) & 0.358 & 0.458 & 0.100 & 0.365 & \textbf{0.007} \\
HN-CP1 (150) & 0.107 & 0.153 & 0.047 & 0.100 & \textbf{0.007} \\
HN-Reserve (72) & 0.111 & 0.139 & 0.028 & 0.097 & \textbf{0.014} \\
\midrule
\multicolumn{6}{c}{\textbf{(b) Sensitivity Envelope Over $\alpha$ in $w_{\mathrm{gap}}(\alpha)$ (8 Independent Cells)}} \\
\midrule
$\alpha$ Weight & 0.00 ($d_{\mathrm{lex}}$ only) & 0.25 & \textbf{0.50 (Frozen)} & 0.75 & 1.00 (Sib. only) \\
\midrule
8-Cell Mean $\Delta$BO (pp) & $+2.35$ & $+2.75$ & $\mathbf{+2.97}$ & $+3.20$ & $+3.16$ \\
Cells Beating All-BO & 6 / 8 & 6 / 8 & \textbf{7 / 8} & 7 / 8 & 6 / 8 \\
\midrule
\multicolumn{6}{c}{\textbf{(c) Transductive Batch-Size Sensitivity (Mean $\pm$ Std HEM, 100 Random Partitions)}} \\
\midrule
Dataset / Backbone (All-BO) & $B=16$ & $B=32$ & $B=64$ & $B=128$ & Full Batch$^{a}$ \\
\midrule
X-Exp-497 / Qwen (0.479) & $0.530 \pm 0.007$ & $0.530 \pm 0.006$ & $0.532 \pm 0.005$ & $\mathbf{0.534} \pm 0.004$ & $0.533$ \\
X-Exp-497 / Phi-4 (0.479) & $0.516 \pm 0.007$ & $0.517 \pm 0.006$ & $\mathbf{0.518} \pm 0.005$ & $0.517 \pm 0.005$ & $0.513$ \\
HN-All-480 / Qwen (0.523) & $0.523 \pm 0.006$ & $0.525 \pm 0.005$ & $\mathbf{0.526} \pm 0.004$ & $0.525 \pm 0.003$ & $0.525$ \\
HN-All-480 / Phi-4 (0.523) & $0.524 \pm 0.005$ & $0.527 \pm 0.004$ & $0.529 \pm 0.004$ & $\mathbf{0.531} \pm 0.003$ & $\mathbf{0.531}$ \\
\bottomrule
\end{tabular}
\par\vspace{0.25em}\raggedright\scriptsize $^{a}$Panel~(c) partitions the combined 222-candidate \texttt{HN-All-480} pool into disjoint batches of size $B$ (hence Full runs a single 222-item batch, whereas \Cref{tab:stage2_blind} strictly concatenates separate CP1 and Reserve decisions).
\end{table}

\begin{table}[h]
\centering
\caption{Pre-annotation prediction freeze inventory and file consistency check for Stage-2 Blind \texttt{Reserve-160} (6 primary artifacts shown out of 12 locked prior to dual annotation; full checksum records are included in the reproducibility package).}
\label{tab:app_freeze}
\footnotesize
\setlength{\tabcolsep}{4.5pt}
\resizebox{\textwidth}{!}{%
\begin{tabular}{llcc}
\toprule
Frozen Prediction Artifact & Model Role & Scope & Consistency Check \\
\midrule
\texttt{adadepth\_m7b\_hn\_reserve\_72\_PRE\_GOLD.csv} & Frozen Qwen2.5-7B \texttt{AdaDepth v1.0} & 72 leaf cands. & Pass \\
\texttt{adadepth\_m4b\_hn\_reserve\_72\_PRE\_GOLD.csv} & Frozen Phi-4-mini \texttt{AdaDepth v1.0} & 72 leaf cands. & Pass \\
\texttt{nli\_deberta\_hn\_all.jsonl} & DeBERTa-v3-large NLI baseline & 222 HN cands. & Pass \\
\texttt{nli\_bart\_hn\_all.jsonl} & BART-large-MNLI baseline & 222 HN cands. & Pass \\
\texttt{bi\_bge\_large\_hn\_all.jsonl} & BGE-large-en-v1.5 Bi-Encoder & 222 HN cands. & Pass \\
\texttt{ce\_minilm\_hn\_all.jsonl} & MiniLM-L6 Cross-Encoder & 222 HN cands. & Pass \\
\bottomrule
\end{tabular}%
}
\end{table}

\FloatBarrier
\bibliographystyle{ACM-Reference-Format}
\bibliography{references}

\end{document}